\documentclass[manuscript,screen]{acmart}
\AtBeginDocument{%
  \providecommand\BibTeX{{%
    \normalfont B\kern-0.5em{\scshape i\kern-0.25em b}\kern-0.8em\TeX}}}

\copyrightyear{2024}
\acmYear{2024}
\setcopyright{rightsretained}
\acmConference[CHI '24]{Proceedings of the CHI Conference on Human Factors in Computing Systems}{May 11--16, 2024}{Honolulu, HI, USA}
\acmBooktitle{Proceedings of the CHI Conference on Human Factors in Computing Systems (CHI '24), May 11--16, 2024, Honolulu, HI, USA}
\acmDOI{10.1145/3613904.3643022}
\acmISBN{979-8-4007-0330-0/24/05}

\usepackage[utf8]{inputenc}
\usepackage{tabularx}
\usepackage{multirow}
\usepackage{enumitem}
\usepackage{color,soul}   
\sethlcolor{white}  
\usepackage{array} 
\newlist{tabitem}{itemize}{1}
\setlist[tabitem]{label=\textbullet, nosep, leftmargin=*,
                  before=\begin{minipage}[t]{\hsize}, 
                  after=\end{minipage}}
\begin{document}

\title[Data Storytelling in Data Visualisation]{Data Storytelling in Data Visualisation: Does it Enhance the Efficiency and Effectiveness of Information Retrieval and Insights Comprehension?}


\author{Hongbo Shao}
\affiliation{%
  \institution{Monash University}
  \city{Melbourne}
  \country{Australia}}
\email{hsha0039@student.monash.edu}

\author{Roberto Martinez-Maldonado}
\orcid{0000-0002-8375-1816}
\affiliation{%
 \institution{Monash University}
 \city{Melbourne}
 \country{Australia}}
\email{Roberto.MartinezMaldonado@monash.edu}

\author{Vanessa Echeverria}
\orcid{0000-0002-2022-9588}
\affiliation{%
 \institution{Monash University}
 \city{Melbourne}
 \country{Australia}}
 \affiliation{%
 \institution{Escuela Superior Politécnica del Litoral}
 \city{Guayaquil}
 \country{Ecuador}}
\email{Vanessa.Echeverria@monash.edu}

\author{Lixiang Yan}
\affiliation{%
 \institution{Monash University}
 \city{Melbourne}
 \country{Australia}}
\email{Lixiang.Yan@monash.edu}

\author{Dragan Gašević}
\affiliation{%
 \institution{Monash University}
 \city{Melbourne}
 \country{Australia}}
\email{Dragan.Gasevic@monash.edu}

\renewcommand{\shortauthors}{Shao, et al.}

\begin{abstract} 

Data storytelling (DS) is rapidly gaining attention as an approach that integrates data, visuals, and narratives to create \textit{data stories} that can help a particular audience to comprehend the key messages underscored by the data with enhanced efficiency and effectiveness. It has been posited that DS can be especially advantageous for audiences with limited visualisation literacy, by presenting the data clearly and concisely. However, empirical studies confirming whether data stories indeed provide these benefits over conventional data visualisations are scarce. To bridge this gap, we conducted a study with 103 participants to determine whether DS indeed improve both efficiency and effectiveness in tasks related to information retrieval and insights comprehension. Our findings suggest that \hl{data stories} do improve the efficiency of comprehension tasks, as well as the effectiveness of comprehension tasks that involve a single insight\hl{, compared with conventional visualisations}. Interestingly, these benefits were not associated with participants' visualisation literacy.


\end{abstract}



\maketitle

\section{Introduction}
The critical role that data plays, and is anticipated to continue playing, in our society is evident through the progressively deeper integration of data-driven decision-making and advanced analytics across various sectors, such as healthcare \cite{sun2023data}, manufacturing \cite{park2023charlie}, education \cite{yang2023pair}, finance \cite{schroeder2020evaluation}, and more. As a result, effective data visualisation has become pivotal in supporting the exploration, analysis, and interpretation of these abundant data \citep{Figueiras2014,lowe2020requirements}. However, not all data visualisations are created with the same intent. Some visualisations are tailored for experts with in-depth knowledge of data analysis. For instance, business analysts might use such visualisations to enhance marketing strategies through a detailed market analysis \citep{Daradkeh2021}. These visualisations are often referred to as \textit{exploratory} visualisations and are primarily aimed at users with expertise in data analysis who are exploring and seeking insights from unfamiliar datasets \cite{Kosara2013,bravo2020immersive,martinez2020data,iliinsky2011designing}.

In contrast, some visualisations are designed to present information to a broader audience, who may not be familiar with specific visualisation techniques but still need to understand the key insights derived from the data \cite{bottinger2020reflections,lee2015people}. For instance, journalists frequently use data visualisations to help the general public understand complex topics such as climate change \citep{aurambout2013simplifying}. Teachers are beginning to employ learning analytics visualisations to gain insights into and enhance their students' performance \citep{Echeverria2017}, while individuals often turn to fitness app visualisations to track their health status \citep{Watson2015}. Such visualisations have an \textit{explanatory} objective, focusing on presenting and communicating insights rather than inviting people to explore the data \cite{Kosara2013,iliinsky2011designing,rodrigues2019once,Echeverria2018}. These explanatory visualisations can be used in various contexts, from illustrating body weight trends in a fitness app \citep[e.g.][]{Murnane2020} to showcasing fluctuating indicators within a company for informed decision-making \citep[e.g.][]{Daradkeh2021}. However, previous empirical studies have shown that the general public often possesses a relatively low level of \textit{visualisation literacy}, meaning they frequently struggle to confidently interpret visual representations of data \citep{maltese2015data,borner2016investigating,lee2015people,Donohoe2020}.

In this context, data storytelling is gaining prominence for helping people grasp key insights from the data \citep{schulz2013towards}. Data storytelling has been defined as \textit{"a structured approach for communicating data insights more effectively to an audience using narrative elements and data visualisations"} (p. 299) \citep[p.299]{dykes2015data}. \hl{While conventional visualisations are often designed to engage the audience in an exploration of the data presented} \citep{Roberts2018explanatory}, \hl{their ability to explain or communicate data insights can be further enhanced by incorporating \textit{data storytelling elements} }\citep{Zdanovic2022}. \hl{These elements emphasise specific messages in the data, for example, by adding text explanations for trends or outliers }\citep{Fan2022deception,ren2023re}, \hl{and by highlighting the results of the analysis in titles }\citep{Kong2018frames} \hl{and text annotations} \citep{stokes2023stricking}. \hl{This approach also involves reducing the emphasis on less relevant visualisation elements and data points }\citep{bach2018narrative,Segel2010,Knaflic2015}.

Various authors have suggested that data storytelling can help in communicating key insights more \textit{effectively} \citep{Segel2010,Knaflic2015,gershon2001storytelling,krum2013cool,Daradkeh2021} and more \textit{efficiently} \citep{Ryan2016, zhang2018converging,zhang2019designing,Zhang2022framework} compared to conventional visualisation. The premise is that this is achieved by reducing the complexity of the information and focusing only on essential data and visual elements. Moreover, \hl{the addition of data storytelling elements to conventional visualisations to create \textit{data stories }}has been seen as a potential bridge to close the comprehension gap by audiences with limited data knowledge or data visualisation literacy \citep{Lee2015more, Echeverria2018, Ma2012}.
Yet, there is limited empirical evidence to back the proposed advantages of data storytelling \hl{elements} in the existing literature. While some studies have explored the potential benefits of data storytelling over conventional visualisations in areas like engagement \citep{boy2015does,zhao2019understanding}, empathy \citep{boy2017showing,morais2021can,Liem2020}, and memory recall \citep{Zdanovic2022}, the results have been mixed. A significant gap remains in empirical studies that address the central premise driving the adoption of data storytelling \hl{in data visualisation}: does it actually assist people in more effectively and efficiently discerning and understanding critical data insights compared to conventional visualisations? Additionally, some preliminary research has examined the potential correlation between audience preferences and personality traits, and data storytelling \citep{aziz2022review}. However, no study has delved into how an individual's visualisation literacy might influence the overall effectiveness and efficiency of data storytelling.

The primary contribution of this paper is the empirical evidence it provides through a comparative study. This study evaluated whether \hl{visualisations with data storytelling elements }improves efficiency (measured by the time taken) and effectiveness (measured by accuracy rates), in two specific tasks: information retrieval (which involves pinpointing key data points) and data comprehension (which entails understanding main insights). Compared against conventional visualisations, these findings can offer actionable insights for visualisation designers, enabling them to make evidence-based decisions on whether to employ data storytelling elements or stick with conventional visualisations for specific tasks, effectively bridging a notable gap in the existing literature. A total of 103 participants, with varied data visualisation experience, visualisation literacy and backgrounds, participated in the study where they were asked to identify relevant data points and key insight(s) from a set of 6 visualisations --3 of which were conventional and the other 3 integrated data storytelling elements. We also assessed participant perceptions of how specific data storytelling elements might influence task completion success rates. Furthermore, we examined the relationship between an individual's visualisation literacy and the effectiveness of data storytelling.

\section{Background and Related Work}
Three areas are relevant to this paper: the theoretical foundations of data storytelling, empirical studies that evaluate its purported benefits, and the role of visualisation literacy. 

\subsection{Foundations of Data Storytelling in Data Visualisation}
\label{sec:foundations}
\hl{Conventional data visualisations primarily focus on presenting raw data in a structured and, ideally, visually appealing manner }\citep{tufte2001visual}. \hl{These visualisations -- such as bar charts, line charts, pie charts, scatter plots, heatmaps, area charts, and bubble charts -- are often intended to serve as tools for users to explore and interpret the data independently }\citep{Roberts2018explanatory}, \hl{offering the advantage of providing a snapshot of the data and laying out facts in a way that can be visually inspected }\citep{Hicks2009perceptual}. \hl{Yet, while conventional visualisations can be valuable, they can sometimes fall short in conveying the deeper narrative behind the data, specially for audiences with little expertise in data analysis }\cite{bravo2020immersive,martinez2020data,iliinsky2011designing,borner2016investigating,maltese2015data}. \hl{Understanding the story behind the data is crucial for gaining meaningful insights and comprehending what the data points or trends in a data visualisation represent in the real world }\citep{Kosara2013}. \hl{Data storytelling is argued to effectively bridge this gap.}

\textit{Storytelling} is the act of weaving facts, insights, emotions and intentions into a narrative that engages audiences and imbues meaning into complex ideas  \citep{Mark17fiction,fog2005storytelling}, often prompting action \citep{dimond2013hollaback}. Applying this definition to the realm of data visualisation, data storytelling can be broadly described as the craft of transforming raw numbers and low-level data points into a cohesive narrative, rendering data not just understandable but actionable. This idea is far from new. \citet{Kosara2013} identified that the early attempts, that resemble today's data storytelling practices, can be traced back to the 18th and 19th centuries when early practitioners of statistics and data analysis demonstrated the power of visually presenting data to show, explain or emphasise the magnitude of certain problems for numerically illiterate decision-makers or general audiences. For example, early \textit{'data stories'} aimed to explain the scope of infectious outbreaks after careful analysis \citep{johnson2006ghost}, present events in a digestible format for educational and training purposes \citep{gershon2001storytelling}, or advocate for social changes \citep{cohen1984florence}. 

In the early 21st century, journalists, equipped with new technologies that made it more accessible to collect, analyse and visualise data, started to use numerical data to support narrative storytelling in various news outlets \cite{showkat2021stories,Segel2010}. As time progressed, professionals across various industries have begun to recognise the importance of communicating data-driven insights in an accessible and compelling manner \cite{riche2018data,Daradkeh2021}. Pioneers in the field of data visualisation, such as Edward Tufte, also emphasised the importance of contextualising data and providing a narrative structure for better comprehension of what the numbers presented in a graph actually mean \citep{tufte2001visual}. In the last few years, several guides on how to do data storytelling have started to emerge \citep{Knaflic2015,feigenbaum2020data,ryan2018visual} making this notion even more attractive and accessible to practitioners across several industries. Nowadays, data storytelling is being used in a wide range of areas such as sports analytics \citep{yu2022basketball,zhi2019gameviews}, education \citep{martinez2020data,wang2019teaching}, psychology \citep{lan2022negative}, social work \citep{Wilkerson2021youth} and heritage transmission \citep{shan2022heritage}. In short, both early and contemporary data storytelling efforts have primarily focused on presenting findings to a broader audience interested in the insights, rather than in the data analysis itself. However, it is only very recently that the concept of data storytelling has begun to be more formally defined and examined.

\citet{Segel2010} were pioneers in establishing the design space of narrative visualisations as a foundation for data storytelling. They described how narrative elements could be integrated with graphics and interaction techniques to organise data for the purpose of conveying a specific message. \citet{Kosara2013} further clarified the term \textit{data storytelling}, arguing that while conventional data visualisation tools can facilitate data analysis, the emphasis of data storytelling is on presenting insights derived from such analysis. To this end, visual cues can be strategically employed to prioritise certain interpretations over others \citep{Hullman2013}. \hl{These cues may include titles }\citep{Kong2018frames,stokes2023stricking} \hl{and graphical or textual annotations }\citep{munzner2014visualization, lai2020automatic, Fan2022deception, Ren2017chartaccent} \hl{that augment existing conventional visualisation components for added clarity or detail or to help guide the reader to understand the data}.

Based on this basic premise, various frameworks for data storytelling have emerged. For example, \citet{bach2018narrative} suggested a set of patterns that could help designers create data stories or analyse the narratives of existing data stories. These patterns are organised into five groups each contributing to the augmentation, flow, framing, emotion and engagement of the data story. \citet{ojo2018patterns} extracted data storytelling patterns from award-winning journalistic data stories, identifying the most common goals of data stories that can be encountered in the media (i.e., to inform, explain, persuade or entertain) and the data visualisation tools that provide the flexibility required for incorporating storytelling elements. \citet{Zhang2022framework} introduced a conceptual workflow demonstrating how raw data \hl{or conventional data visualisations} could be transformed into data stories that emphasise specific elements. \citet{Echeverria2017} presented a model and implementation for automating the creation of data stories based on pre-set parameters by end-users. These advancements indicate a growing interest in formalising the design characteristics of data 
storytelling.

Despite the identification of numerous patterns and design elements in academic literature, most current data storytelling studies converge on \hl{a limited set of key principles for data storytelling}, irrespective of the diverse ways it can be implemented. \citet{Echeverria2017} \hl{synthesised five principles for transitioning from conventional visualisations to data stories that i) direct the user's attention to ensure visual guidance, and ii) facilitate the narrative to deliver the intended message or data insight(s)} \citep{Zdanovic2022}. \hl{These principles are based on the work of }\citet{Ryan2016} and \citet{Knaflic2015}, \hl{and they align with the process proposed in the data storytelling framework by }\citet{Zdanovic2022} \hl{which we use in this paper to enhance conventional visualisations with data storytelling elements (details are provided in the next section)}:

\begin{itemize}
\item \textbf{Principle \#1} -- \textbf{Identifying an explicit goal:} This involves understanding the \textit{audience}, their level of technological savviness, and subject matter expertise, a common theme across data storytelling literature \citep{bach2018narrative,Segel2010,Lee2015more,dykes2015data}.

\item \textbf{Principle \#2} -- \textbf{Removing redundant elements:} This involves eliminating elements that offer no additional value, such as unnecessary data labels, markers, grids, legends, tick marks, and axis labels. This principle is foundational in effective data visualisation \citep{tufte2001visual} and is emphasised in data storytelling guides as a crucial decluttering step \citep{Knaflic2015,feigenbaum2020data,ryan2018visual}.

\item \textbf{Principle \#3} -- \textbf{Employing storytelling elements with discernment:} This is, using annotations and visual cues judiciously to clarify or to group related data. Yet, an excess of annotations can clutter the visualisation and distract the audience \citep{kong2019understanding}, potentially counteracting the goal of steering attention  \citep{kalyuga2009expertise}.

\item \textbf{Principle \#4} -- \textbf{Capturing interest:} This involves minimising irrelevant visual elements (i.e., declutter) while emphasising crucial components, for example, by using colours or bold/thick elements for highlights \citep{Knaflic2015,feigenbaum2020data}.

\item \textbf{Principle \#5} -- \textbf{Calling for action:} This involves creating a coherent and succinct visualisation that conveys the essential narrative, for example, by guiding users toward the intended action through an explicit title derived from the data \citep{bach2018narrative,ojo2018patterns}.
\end{itemize}

\subsection{Benefits of Data Storytelling}
Various authors have suggested that \hl{visualisations that contain data storytelling elements, commonly referred to as 'data stories',} can help in communicating key insights more \textit{effectively}. For instance, \citet{Segel2010} posited that integrating narratives with graphics offers a richer medium for communicating analysis results than merely using regular graphical tools for data analysis. Similarly, \citet{gershon2001storytelling} argued that incorporating storytelling elements into conventional data visualisations can contribute to the communication of information effectively and intuitively. \citet{Daradkeh2021} extended this argument, suggesting that such an approach can lead to improved decision-making and business performance. Authors like \citet{Knaflic2015} and \citet{krum2013cool} have explained that \hl{visualisations enhanced with data storytelling elements} offer an advantage in that they leverage humans' efficient perceptual skills for visual pattern recognition. Indeed, the concept of pre-attentive processing -- which is the subconscious stage that enables people to quickly identify visual elements like colour, shape, and orientation -- has provided foundational principles for the field of information visualisation design \citep{Hicks2009perceptual}. Additionally, other researchers \citep{Ryan2016, zhang2018converging, zhang2019designing, Zhang2022framework} have highlighted data storytelling's potential for \textit{efficient} communication. They argue that \hl{visualisations enhanced with data storytelling elements} can outperform conventional visualisation techniques when the design concentrates exclusively on essential data and visual elements, thereby reducing complexity and, consequently, shortening the time needed to comprehend the data.
  
Despite the above claims in the literature, empirical evidence supporting the relative effectiveness and efficiency of data storytelling over conventional visualisations remains limited. Some research has empirical examined other potential advantages; for example, \citet{zhao2019understanding} found that caption-enriched visualisations are more engaging and easier to use than those without. Similarly, the qualitative analysis from the same study suggested that enhancements in data storytelling aid focus and memory. However, a controlled study by \citet{Zdanovic2022} found no significant impact of \hl{visualisations enhanced with data storytelling elements} on either short-term or long-term memory recall when compared to conventional visualisations.

Mixed results also emerged in studies investigating data storytelling's potential for fostering empathy or influencing audience attitudes. For instance, controlled studies by \citet{boy2017showing} and \citet{morais2021can} did not find evidence supporting the idea that adding anthropomorphic elements to visualisations enhances empathy. Likewise, \citet{Liem2020} found no significant difference in empathetic responses between data stories using personal narratives and conventional exploratory visualisations.

An important gap remains unaddressed in the literature: it is unclear whether \hl{adding data storytelling elements to conventional visualisations} aid in the effective and efficient identification of insights and corresponding critical data points when compared to conventional visualisations. While small-scale eye-tracking studies \citep{de2014evaluating, Echeverria2018, Pozdniakov2023how} suggest that data storytelling elements can attract users' attention, they do not directly examine whether such elements facilitate better or faster comprehension of the data.

\subsection{Visualisation Literacy}
The purpose and message of a data visualisation may not always be easily and accurately understood, even if the visualisation is well-designed \citep{Donohoe2020}. As mentioned above, initial attempts at data storytelling aimed to bridge this comprehension gap for audiences unfamiliar with data and visualisations \citep{Kosara2013}. Today, data storytelling often targets groups without formal training in data or visualisation, such as teachers \cite{Pozdniakov2023how}, healthcare practitioners \citep{fernandez2021storytelling}, and the general public \citep{Ma2012}. The construct that encapsulates the skills needed to effectively interact with data visualisations is termed "visualisation literacy" (VL) \citep{Lee2017}. 

\textit{Visualisation literacy} refers to the ability to accurately interpret visual representations of data and to efficiently and confidently extract, process, and draw conclusions from such visualisations \citep{Boy2014, firat_intliteracy_2022}. Intriguingly, multiple studies have shown that the general public often struggles with interpreting visual data representations, indicating a widespread low level of VL \citep{maltese2015data, borner2016investigating, lee2015people, Donohoe2020}. For instance, a study by \citet{maltese2015data} involving higher education students revealed that most had difficulty understanding even basic data visualisations, such as simple bar charts. \citet{Borner2016} found similar results in a study involving museum attendees. Likewise, many teachers find it challenging to interpret educational dashboards, primarily due to a lack of relevant data skills \citep{ndukwe2020teaching}.

When evaluating the effectiveness and efficiency of data stories, it is crucial to consider the potential impact of an individual's level of VL. To date, the only study that has examined the intersection of VL and data storytelling is by \citet{Pozdniakov2023ectel}. This small-scale eye-tracking study involved 23 teachers with varying levels of VL who were tasked with reviewing visualisations presented in dashboards. Some were enriched with data storytelling elements (i.e., emphasis on certain data points) while others maintained a control version containing conventional visualisations. The authors found that teachers with lower VL levels potentially benefit the most from data storytelling enhancements in terms of reduced cognitive load. Yet, the relationship between individual visualisation literacy levels and the potential benefits of data storytelling in supporting effective and efficient insight identification remains largely unexplored.

\subsection{Contribution to HCI and Research Questions}

To address the gaps identified in the existing literature, this paper presents a study aimed at providing empirical evidence on the efficiency and effectiveness of data storytelling compared to conventional visualisations. Specifically, we aimed to quantitatively investigate whether visualisations enhanced with data storytelling elements reduce the time required to answer questions based on visualisations and improve accuracy in tasks related to information retrieval (i.e., identifying key data points that support a given insight) and comprehension (i.e., understanding the actual insight or insights). Additionally, we qualitatively examined participants' perceptions regarding the helpfulness of data storytelling elements to potentially elucidate the quantitative results. Finally, we assessed the influence of individuals' visualisation literacy on the effectiveness of data storytelling \hl{elements}. As a result, we proposed the following research questions:

\begin{itemize}
\item \emph{\textbf{RQ1:} To what extent does \hl{data stories} assist individuals in discerning and understanding critical data insights more \textit{\textbf{efficiently}} compared to conventional visualisations?}
\item \emph{\textbf{RQ2:} To what extent does \hl{data stories} assist individuals in discerning and understanding critical data insights more \textit{\textbf{effectively}} compared to conventional visualisations?}
\item \emph{\textbf{RQ3:} To what extent do the potential improvements in efficiency and effectiveness offered by data storytelling \hl{elements} vary between \textbf{information retrieval and comprehension} tasks?}
\item \emph{\textbf{RQ4:} What are participant’s perceptions on the helpfulness of \textbf{data storytelling elements} (e.g., annotations, visual highlights, and titles -- see details in the next section)?}
\item \emph{\textbf{RQ5:} To what extent do the relationships between individuals' \textbf{visualisation literacy} and their task efficiency and effectiveness vary between data storytelling and conventional visualisation conditions?}
\end{itemize}

\section{Method}
This section presents the study's components, covering:  i) the data storytelling design elements, ii) dataset and materials used, iii) the study procedure, iv) the participants, and v) the analysis.

\subsection{Data Storytelling Design Elements and Transformation Process}
\label{sec:elements}
To reliably compare conventional visualisations with data stories -- those enhanced with data storytelling elements -- we first needed to clearly identify these data storytelling elements. 
\hl{To maintain alignment with prior research evaluating visualisations enhanced with data storytelling elements, we established a set of design actions that involve adding or eliminating visual design elements to and from conventional visualisations to transform them into a data story. In doing so, we built on the framework for data storytelling proposed by }\citet{Zdanovic2022}, \hl{which is itself grounded in the data storytelling principles put forth by }\citet{Echeverria2017}, \citet{Ryan2016}, and \citet{Knaflic2015} \hl{as introduced in Section }\ref{sec:foundations}. \citet{Zdanovic2022} \hl{describe a three-phase process comprising i) initially exploring the data to find the insight(s) to be communicated, ii) crafting the visualisation, and iii) telling the story. The creation of a conventional visualisation typically involves phases i) and ii). In contrast, creating a data story entails considering the following data storytelling elements in the order presented below, particularly in phases ii) and iii). Notably, we excluded interactive elements from our study to prevent the design decisions from being complicated by confounding variables and to ensure consistency with previous data storytelling evaluations }\citep{Zdanovic2022, zhao2019understanding, boy2017showing, morais2021can}. \hl{Once the insights are identified, we considered the following data storytelling design elements and concrete design actions for crafting the data stories from conventional visualisations, adhering to the principles introduced in Section }\ref{sec:foundations}, \hl{and are as follows:}

\begin{figure*}[h!]
\centering
  \includegraphics[width=1\textwidth]{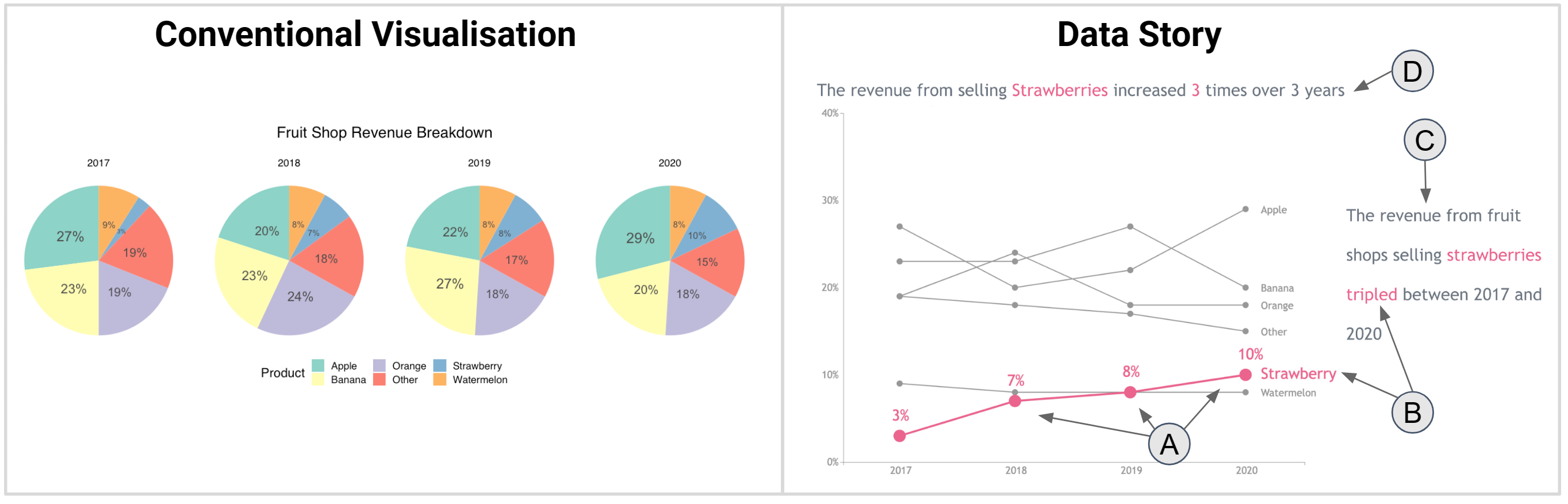}
  \caption{Illustrative example: Conventional data visualisation (left) converted into a data story (right). The latter features several data storytelling elements: A) \hl{annotated} data points; \hl{B) colour emphasis} in text and lines; C) textual annotations; and D) an explanatory title that summarises the key insights. }
  \label{fig:design_elements}
\end{figure*}

\textbf{\hl{1-- }Using the right visualisation technique according to the storytelling goal} (\hl{in phase ii -- }principle \#1, in Section \ref{sec:foundations}). Selecting the appropriate visualisation technique, such as a bar chart or scatter plot, according to the data communication goal is crucial for effectively conveying the relevant data insight(s) and setting the stage for a compelling story \cite{tufte2001visual,few2004show}. \hl{Adhering to } \citeauthor{Knaflic2015}'s \citep{Knaflic2015}\hl{ recommendations, effective data storytelling may, although not always necessary, involve changing chart types to better convey the narrative. For example, to highlight temporal trends, switching from pie charts to a line chart can be more illustrative, as shown in Figure} \ref{fig:design_elements}. 

\textbf{\hl{2-- }Eliminating clutter} (\hl{in phase ii -- }principles \#2 \& \#3). Reducing extraneous visual elements can make a visualisation more easily interpretable, allowing viewers to focus only on key elements \citep{tufte2001visual,Heer2012interactive}. While this is a principle \hl{that is also generally applicable to any conventional data visualisation} (i.e., reducing the data-ink ratio \citep{tufte2001visual}), it becomes vital in data storytelling to allow other visual elements to stand out and effectively draw attention. This is illustrated in Figure \ref{fig:design_elements} (right) by minimising axis information, removing details for less relevant data points, labelling data series directly instead of using legends, and using minimal colour variation. Importantly, decluttering can also be applied to the excessive use of annotations and other data storytelling elements (see below). 

\textbf{\hl{3-- }Directing attention by emphasising data points and trends} (\hl{in phase iii -- }see principle \#4). \hl{For the specific case of data stories, the counterpart to decluttering is adding emphasis. This can be achieved through techniques like employing contrasting colours, highlighting specific data points, or making certain trend lines bolder. These techniques aid in directing the viewer's attention towards crucial data points and trends associated with the main insights identified in phase i, enabling them to comprehend the story being conveyed} \citep{Knaflic2015}. \citet{ware2019information} explained how using colour could guide the viewer's attention, while \citet{few2004show} emphasised the importance of effectively using visual hierarchy to direct attention. For instance, in Figure \ref{fig:design_elements} (right), this is illustrated by emphasising the data series corresponding to strawberry revenue with a thick, coloured line, including details/annotations for relevant data points (A), and consistently using the same colour in the text, lines and other parts of the visualisation to refer to these data points (B).

\textbf{\hl{4-- }Adding annotations} (\hl{in phase iii -- } principles \#3 \& \#4). Annotations provide contextual or explanatory information directly on the visualisation, helping to clarify or further emphasise key points and trends \citep{stokes2023stricking,Hullman2013}. As previously mentioned, annotations can be graphical or textual. In our study, following \citet{Knaflic2015}'s recommendations, \hl{we mainly utilised textual annotations for two purposes:  to \textit{explain} data points or trends that are relevant to the insight(s) identified in phase i, and to convey insights in straightforward language. An example of this is depicted in Figure }\ref{fig:design_elements} (right-C), \hl{which demonstrates how the narrative succinctly explains the primary insight -- the tripling of strawberry sales revenue -- using plain language for clarity}.

\textbf{\hl{5-- }Adding an explanatory title} (\hl{in phase iii -- } principle \#5). A well-crafted title not only describes the visualisation but also guides the viewer towards the intended conclusion or call to action, serving as a headline for the data story. \citet{Knaflic2015} recommends adding a prescriptive title that summarises key findings or even suggests a course of action as a powerful element in data storytelling. \citet{Pozdniakov2023how} and \citet{Kong2018frames} demonstrated that adding a title on top of each chart drives attention and can potentially guide interpretation. An example of a title that briefly summarises a key insight is shown in Figure \ref{fig:design_elements} (right-D). \hl{In our study, we added an explanatory title to each data story to consistently summarise the insight(s) identified in phase i.} 

%

\subsection{Dataset and Materials}
\subsubsection{Dataset}
For this study, we opted for a topic of broad interest: climate change and global warming, which are increasingly significant global issues. Some associated data can be challenging for the general public to comprehend without advanced visualisation skills \citep{Sheppard2005}. We chose to use open-source data from the "\textit{Our World in Data}" (2020) website\footnote{\url{https://ourworldindata.org/}} to help us determine if data storytelling could be beneficial in situations where participants' understanding of the visualisation is crucial \citep{Sheppard2005}.

\subsubsection{Tools}
All visualisations without data storytelling elements (referred to as conventional data visualisations) were generated using Tableau software\footnote{\url{https://www.tableau.com/}} \citep{Batt2020,Hoelscher2018}. In contrast, visualisations featuring data storytelling elements (referred to as data stories) were modified using LucidChart\footnote{\url{https://www.lucidchart.com/}}, a free graphic design tool, to incorporate these elements. The design of the visualisations was jointly agreed upon by all the co-authors of the paper in multiple iterations, making sure the data storytelling elements above were applied. Details about the specific visualisations used in the study are provided below. 

\subsection{Study Design and Procedure}

We designed a controlled study in which we manipulated a single variable: visualisation type (the conventional visualisation condition, or CV, versus the data storytelling condition, or DS). We opted for a within-subjects study design, wherein the same participant would be presented with both conventional visualisations and data stories, with the order of presentation counterbalanced through randomisation. That is, each participant served as their own control, enhancing the statistical power and sensitivity in detecting effects as noted in \cite{cohen2013statistical}. This approach is particularly important, considering we also incorporated individual's visualisation literacy into the analysis of RQ5.

The study was set up as a survey using Qualtrics\footnote{\url{https://www.qualtrics.com/}}, an online platform for designing and conducting experiments. We utilised the Qualtrics service hosted by the authors' institution. Participant recruitment was conducted via Prolific\footnote{\url{https://www.prolific.co/}}, an online platform that supports academic research. Both services could be integrated so that the participants recruited on Prolific could be redirected to the corresponding online experiment on Qualtrics. This allowed for data verification before confirming the participants' valid participation, enhancing the reliability of their responses. 

\begin{figure*}[h!]
\centering
  \includegraphics[width=.99\textwidth]{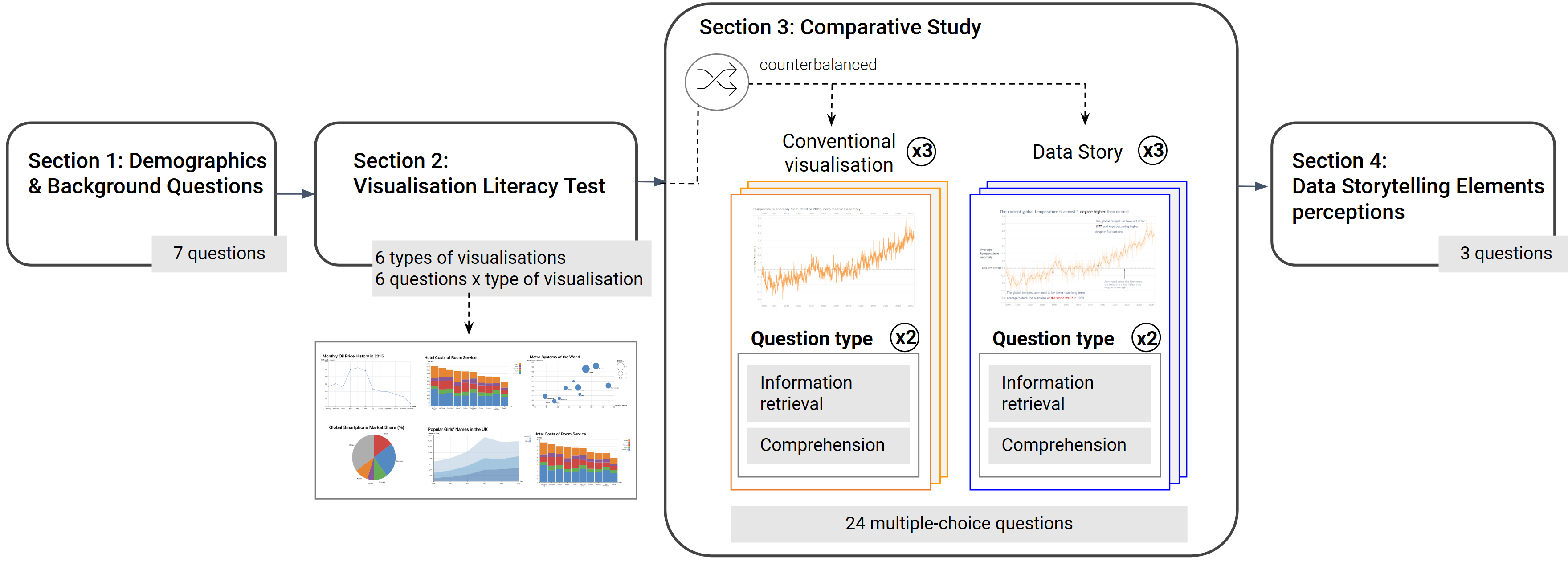}
    \setlength{\belowcaptionskip}{-5pt}
  \caption{The structure of the study. Section 1 included a set of demographic and background questions. Section 2 included a visualisation literacy test adapted to the visualisation techniques utilised in our study \cite{Lee2017}. Section 3 was the main comparative study in which a set of multiple-choice questions were posed to participants based on 3 conventional visualisations and 3 data stories (counterbalanced order); questions were of two types -- information retrieval and comprehension. Section 4 included questions related to participants' perceptions of the data storytelling visual elements they used while answering the questions in Section 3. } 
  \label{fig:study}
\end{figure*}

\hl{The study was structured to be completed within 40 minutes, and participants received AUD \$14 as compensation for their time. Additionally, although participants were compensated for 40 minutes of their time, they were encouraged to complete the study "as fast as possible" for efficient earning.} The study consisted of four sections: i) an introduction, including demographic and background questions; ii) a visualisation literacy test (addressing RQ5); iii) the main comparative study (RQ1, RQ2, RQ3, and RQ5); and iv) questions concerning perceptions of data storytelling elements (RQ4). Ethics approval was obtained from Anonymous University (Project ID: Anonymised). Participants were asked to consent before participating in the study. Figure \ref{fig:study} illustrates the study procedure and maps each section to our research questions (RQs). Details are presented below (further details are presented in the supplementary material of this paper for replication purposes). 

\subsubsection{Section 1 -- Demographics and background questions}
First, participants were asked demographic questions, namely: i) self-reported gender, ii) age, iii) highest level of education, iv) region, v) career, and vi) whether their career is related to data analysis. We then inquired about participants' self-assessed experience in data analysis and data visualisation (e.g., "choose the circumstance that best describes your experience with data analysis"). To enhance the response rate for demographic and background questions, we situated them in the initial section of the survey (Figure \ref{fig:study}-Section 1).

\subsubsection{Section 2 -- Visualisation literacy test}
The objective of this section was to estimate each participant's visualisation literacy (VL) to address RQ5 (see Figure \ref{fig:study}-Section 2). To estimate a participant's VL pertinent to this study, we employed a modified version of the Visualisation Literacy Assessment Test (VLAT), developed by \citet{Lee2017}. The original test covers 12 types of visualisations, each with 6 questions. However, some of these were not applicable to the visualisations featured in our study (e.g., a map of the United States).  To avoid participant fatigue and focus on the specific visualisation types used in our study -- line charts, bar charts, stacked bar charts, pie charts, stacked area charts, and bubble charts -- we selected only the VLAT questions relevant to these visualisation types. This resulted in six sets of questions, each related to one of the six visualisation types. This approach ensured that the assessment measured participants' abilities in the context of the visualisations they encountered later. We employed the corrected score of the VLAT to neutralise the impact of guessing, as suggested by \citet{Frary1988} and \citet{Lee2017}.

\subsubsection{Section 3 -- The comparative study}
\hfill \break

\begin{figure*}[h!]
\centering
  \includegraphics[width=1\textwidth]{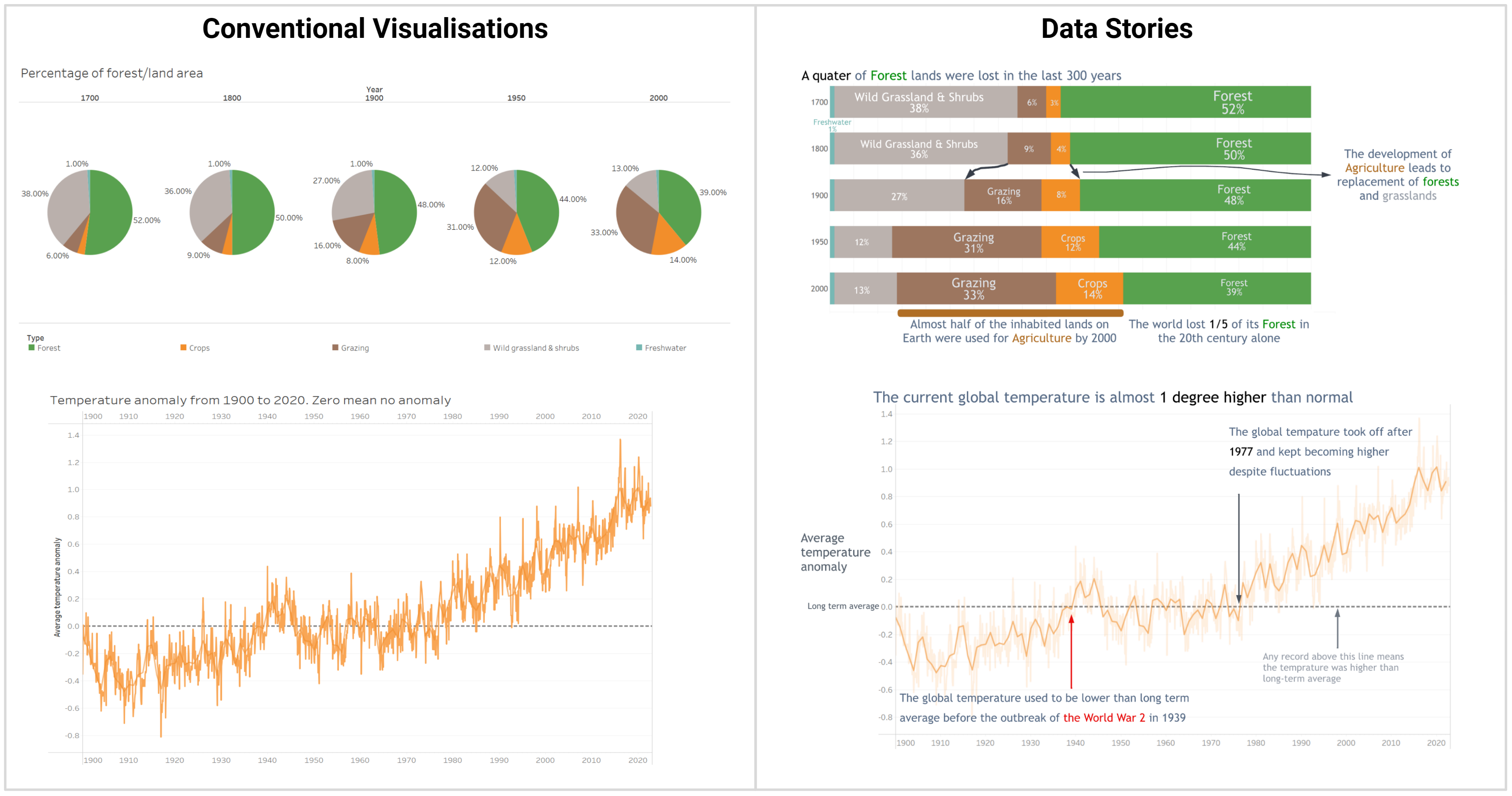}
  \caption{Examples of two pairs of visualisations shown to participants: conventional visualisations (left) and their data storytelling versions (right)} 
  \label{fig:pair}
\end{figure*}

\textbf{Conventional visualisations versus data stories}. This consisted of the core questions of this survey (see Figure \ref{fig:study}-Section 3). We designed 6 pairs of visualisations, each showcasing different data but conveying the same insight within a pair. Specifically, we created six conventional visualisations (cv1-6) alongside six alternate versions that applied data storytelling principles (ds1-6). \hl{These data stories were created following the transformation process to convert conventional visualisations to those with data storytelling elements, as proposed by }\citep{Zdanovic2022} \hl{and detailed in Section }\ref{sec:elements}. \hl{
Two researchers participated in several discussion sessions to collaboratively identify the insights from each visualisation based on the data and metadata provided in the \textit{Our World in Data} repository (phase i) to then apply the transformation process (phases ii and iii), ensuring that aspects such as the type of chart, decluttering actions, emphasis techniques, annotations, and explanatory titles all effectively contributed to conveying the insights. Later, a third researcher validated the insights and the data stories, and further participated in discussion sessions with the other researchers to agree on the final set of visualisations. All visualisations included the complete range of data storytelling elements discussed above. Two example conventional visualisation--data story pairs (cv1--ds1 and cv2--ds2) are} shown in Figure \ref{fig:pair}. To facilitate a balanced exposure to both conventional and data storytelling visualisations, each participant was randomly presented with three items from each condition, and in a random sequence. For instance, one participant might be shown visualisations cv1, cv2, cv3 paired with ds4, ds5, ds6, while another could encounter cv4, cv5, cv6 and ds1, ds2, ds3. This method not only ensured an even distribution of both conventional visualisations and data stories but also helped mitigate any learning or practice effects. No explicit indication was provided to participants regarding which visualisation was conventional and which was a data story.

The data visualisations were designed to be diverse in terms of both visualisation complexity and frequency of use across varied audiences. To achieve this, and as described in the previous section, we included a range of chart types: line charts, bar charts, pie charts, stacked bar charts, stacked area charts, and bubble charts. According to \citet{Segel2010}, line charts, bar charts, and pie charts are the most commonly used visualisation techniques and are relatively easy to understand \citep{Lee2017}. Stacked bar charts and stacked area charts have been found to be slightly more challenging for participants, as indicated by the difficulty scores calculated by \citet{Lee2017}, yet they are also commonly used. Bubble charts have been considered more difficult to interpret, according to their difficulty scores, and are less commonly used \citep{Lee2017}. This assortment of chart types provided a comprehensive range of visualisations (see details in Table \ref{tab:charts}). In our study, the distribution mirrored the commonality of these charts: we included three chart types that are most common, two that are relatively common, and one that is less commonly used. The choice of chart for the conventional visualisations was also influenced by the specific data being displayed. We intentionally designed the conventional mirroring of the original data source made public through \textit{Our World in Data}.

\renewcommand{\arraystretch}{3}
\begin{table}[h]
\centering
\caption{Overview of the pairs of visualisations with and without data storytelling elements. The symbol "*" represents the pairs in which the chart type changed from conventional visualisation to the data storytelling condition. The actual visualisations can be found \href{https://anonymous.4open.science/r/data-storytelling-chi-2024-F1BF/SurveyVisualisations.pdf}{here}.}
\label{tab:charts}
\begin{tabular}{|p{3.5cm}|p{4cm}|p{6cm}|}
\hline
Conventional visualisation & Data storytelling  & Questions types \\
\hline
Line chart & Line chart & \begin{minipage}{6cm} 2x information retrieval questions \\ 2x comprehension questions (multiple insights) \end{minipage}\\
\hline
Pie chart & Line chart * & \begin{minipage}{6cm} 2x information retrieval questions \\ 2x comprehension questions (single insight) \end{minipage}\\
\hline
Pie chart & Stacked bar * & \begin{minipage}{6cm} 2x information retrieval questions \\ 2x comprehension questions (multiple insights) \end{minipage}\\
\hline
Stacked area chart & Area chart & \begin{minipage}{6cm} 2x information retrieval questions \\ 1x comprehension questions (single insight) \\ 1x comprehension question (multiple insights) \end{minipage}\\
\hline
Stacked area chart & Line chart * & \begin{minipage}{6cm} 2x information retrieval questions \\ 1x comprehension questions (single insight) \\ 1x comprehension question (multiple insights) \end{minipage}\\
\hline
Bubble chart & Bubble chart & \begin{minipage}{6cm} 2x information retrieval questions \\ 2x comprehension questions (single insight) \end{minipage}\\
\hline
\end{tabular}
\end{table}
\renewcommand{\arraystretch}{1}

We included text annotations, data point annotations, colour emphasis, and explanatory titles in all visualisations with data storytelling elements. Moreover, since the change of chart type can significantly affect visualisation and different chart types require distinct visualisation skills from participants \citep{Lee2017}, we limited the change of chart types to half of the visualisation sets (3 sets) (see details in Table \ref{tab:charts}, columns 1 and 2). 

\textbf{Types of questions}. 
Each time a participant was presented with a visualisation, they were asked to respond to four multiple-choice questions. Each question was categorical, and the correct response was only one out of four predefined options. For each pair of visualisations, the same set of questions was asked consistently for both versions --the conventional visualisation and the data story -- to different participants. 

We defined these questions based on Bloom's taxonomy \citep{bloom1984bloom}. 
Bloom's taxonomy has been extensively used to categorise questions into varying levels of complexity (i.e., knowledge, comprehension, application, analysis, synthesis, and evaluation). It has also been proposed for assessing the kinds of tasks that data visualisation can enable \citep{arneson2018visual,Burns2020}, as well as for supporting the development of visualisation skills \citep{mnguni2016assessment,Byrd20197d}. 
\hl{ Following the guidelines proposed by} \citet{Burns2020}, \hl{we designed two types of questions based on the first two levels (knowledge and comprehension) of Bloom's taxonomy. We selected these two levels because they align with the goal of this research (RQ3). Further levels require other types of interventions and study designs that go beyond the purpose of the current study.} In the context of information visualisation, \hl{knowledge (Level 1) }questions in Bloom's taxonomy serve to assess whether users can identify and select relevant data points, categories, or trends from a data set. This aligns with our goal to assess whether data storytelling supports \hl{\textbf{\textit{information retrieval}} tasks}. 
Comprehension \hl{questions (Level 2) }relate to questions that assess the ability to understand, interpret, and derive meaningful insights from the data. Data comprehension, in this sense, goes beyond mere retrieval and involves interpretation, assessment, and decision-making. It involves \hl{identifying various data points, making comparisons,} and understanding the underlying meaning, \hl{which is aligned with the \textit{\textbf{comprehension}} task}. Questions of this type were subdivided into two sub-types. The first sub-type asked participants about a \textit{single insight} that the data visualisation aimed to provide and allowed the participants to choose the correct answer. 
The second sub-type requires participants to gain a comprehensive understanding of \textit{multiple insights}. The question presents four statements, and the participants were required to identify the invalid statement.

\hl{For each pair of visualisations, we added two information retrieval (Level 1) questions and two comprehension (Level 2) questions (see Table} \ref{tab:charts},\hl{ column 3).
Table} \ref{tab:questions} \hl{(row 1) shows an example of an information retrieval question that requires the participants to retrieve a single data point from the visualisation and choose the correct value among the response options }\citep{Anderson2001}.
Table \ref{tab:questions} \hl{(row 2) shows an example of a comprehension question that asks the participants to identify a single insight by comparing various data points. }
The comprehension question sub-types were equally distributed across all pairs of visualisations and depended on the number of insights that could be extracted from the data shown in each visualisation pair (see Table \ref{tab:charts}, column 3). Table \ref{tab:questions} (row 3) \hl{presents an example in which a participant needs to identify the incorrect insight among the other three correct insights communicated in the conventional visualisation or the data story. 
Two researchers collaboratively designed these questions independently of the type of visualisation (conventional or data story). A third researcher validated these questions. Discussion sessions were conducted among the three researchers to reach a consensus on the final set of questions. They ensured there was an alignment between the questions asked and the information highlighted in the visualisations with data storytelling elements. }

The time taken to complete each question was recorded to assess the efficiency of data storytelling (RQ1) \citep{Zhu2007,Garlandini2009}, while the correctness rate of responses was recorded to shed light on its effectiveness (RQ2) \citep{Zhu2007,Garlandini2009}.

\begin{table}[h]
\caption{Examples of information retrieval and comprehension questions asked for both conventional visualisations and data stories}
\label{tab:questions}
\centering
\begin{tabularx}{\textwidth}{|l|l|>{\raggedright\arraybackslash}X|}
\hline
\textbf{Bloom's taxonomy} & \textbf{Question type} & \textbf{Examples} \\
\hline
Level 1: Knowledge & Information Retrieval (x2) & What proportion of land was used for grazing in 1900?
\begin{tabitem}
 \item [1.] 31\%
 \item [2.] 27\%
 \item [3.] 16\%
 \item [4.] 8\%
\end{tabitem} \\
\hline
\multirow{2}{*}{Level 2: Comprehension} & Comprehension (single insight) & Emissions from which fuel/industry increased most significantly between 1950-1980? 
\begin{tabitem}
 \item [1.] Gas
 \item [2.] Oil
 \item [3.] Coal
 \item [4.] Flaring
\end{tabitem} \\
\cline{2-3} & Comprehension (multiple  insights) & Choose the incorrect statement:
\begin{tabitem}
 \item [1.] More than half of the usable lands on Earth are now used for agriculture
 \item [2.] The percentage of Wild Grassland \& Shrubs increased in the recent 50 years
 \item [3.] The development of agriculture between 1800 and 1900 greatly reduced the percentage of Wild Grassland \& Shrubs and Forest
 \item [4.] $\frac{1}{5}$ of Forest was lost in the 20th century alone
\end{tabitem} \\
\hline
\end{tabularx}
\end{table}

\subsubsection{Section 4 -- Data storytelling elements questions}

In this section, we revealed to participants that three earlier visualisations contained data storytelling elements. We clarified this by displaying a side-by-side comparison of a conventional visualisation and a data story, both illustrated in Figure \ref{fig:design_elements}. We annotated each storytelling element within the visualisations to help participants associate the terms with specific visual cues. The highlighted elements were: a) annotated data points, b) colour/thickness emphasis, c) text annotations, and d) an explanatory title. Participants then shared their preferences on the most useful -- or not useful at all -- elements for extracting information, with the option to further explain their choices through an open-ended question.


\subsection{Participants}

We aimed to recruit 100 participants, based on the participant numbers from previous similar data visualisation and storytelling studies \citep{Hullman2013,Robinson2014,Zdanovic2022}, requiring them to have advanced English skills and to use a laptop or desktop for consistent visualization displays. We conducted two pilot studies before rolling out the full study, involving three university students to fine-tune the survey, with the second pilot’s data being incorporated (2 participants). Participants were allowed to skip up to three questions but skipping more than three or not completing the entire survey invalidated their responses. We received 110 responses, of which 9 were invalid due to incompleteness, resulting in a total of 103 valid responses, which includes the 2 participants from the second pilot.

The participants came from six different regions, with the majority coming from North America/Central America (39), Europe (34), and Africa (22). The remainder were from South America (4), Oceania (3), and Asia (1). They identified as female (42), male (60), and non-binary (1), averaging 26.6 years of age (std. dev.= 6.8). Most attained a Bachelor's degree (48) or high school education (37), while others held Master's (13) or PhDs (2). A total of 36 participants were students at the time of the survey. Examining the participants' fields of employment, a significant portion worked in the Technology and IT sector (12), pursued by roles in sales, retail, and customer service (7), and administration and management (6). Seven participants were actively seeking employment or were currently unemployed. A few were job-seeking (7) or represented other sectors, including healthcare and engineering. 

Around a third of participants (32) engaged in data analysis in their jobs. About half (51) had a basic understanding of the concept, while 30 were only familiar with the term. A small group had substantial knowledge (15) or professional expertise (3), whereas 4 had no exposure to the domain. Concerning data visualization, most encountered it through newspapers, apps, and websites (48), and in their professional roles (36), although just a few considered themselves professionals in this field (5). This means that, overall, most participants lacked expertise in data analysis and visualization.

\subsection{Analysis}

\subsubsection{Metrics}
We defined and calculated the following metrics to answer our research questions. All those metrics were calculated both for data storytelling and conventional data visualisation: 
\begin{itemize}
\item Correct Rate: In the main comparative study detailed in section 3, each question regarding individual pairs of visualisations carried a raw score of 1 for a correct response. We separately calculated the total raw scores for questions grounded on conventional visualisations and those focused on data storytelling, recording the number of correct answers given by participants in each category. To adjust for the possibility of correct answers attained through guessing, we applied the "correction for guessing" formula proposed by \citet{Frary1988}. This formula is commonly used in the evaluation of multiple-choice questions and aims to offer a more precise measure by mitigating the impact of random guessing:

\begin{equation}
CS = \text{Right} - \text{Wrong} / (\textit{k} - 1)
\end{equation}
where CS is the \textit{correct rate} of a participant (corrected for guessing), \textit{Right} denotes the participant's number of correct answers (i.e., the raw score), \textit{Wrong} signifies the number of incorrectly answered questions, and \textit{k} represents the number of choices per question, which in our study is four. The range for this metric is 0 to 1.
\item Average Success Time: The average success time was calculated by dividing the sum of the time intervals -- measured from the moment a question was presented to participants (page opened) until the time the participant submitted a correct answer (page closed and the answer was correct) -- by the total number of correct answers. This metric was calculated separately for each participant, also distinguishing between questions based on conventional visualisations and those based on data storytelling.

\item Visualisation Literacy Score: Each question in the contextualised VL test that we applied in our study carried a raw score of 1 for a correct answer. The total raw score was calculated based on the number of questions the participants answered correctly, divided by the total number of VL questions. To adjust for the influence of correct answers potentially attained through guessing, we also applied the "correction for guessing" formula \citep{Frary1988}. The range for this metric is 0 to 1.
   \end{itemize}

\subsubsection{Normality consideration}
 We employed the Shapiro-Wilk W-test to assess the normality of the correct rate, average completion time, and visualisation literacy scores \citep{Royston1992}. Our analysis provided strong evidence that these variables were not normally distributed. 

\subsubsection{Analysis for each research question}
To address RQ1, we utilised R to compute the median and interquartile range (IQR) of the \textit{average success time }metric, visualising the results with a box plot. Subsequently, we applied the Wilcoxon signed-rank test, suitable for paired data and capable of accounting for individual differences by using each participant as their own control, to evaluate whether the success time metric under data storytelling significantly differed from those observed with conventional data visualisations for the same participants. We adopted a p-value threshold of less than 0.05 to determine significance as guided by \citep{Nachar2008, Tandon2023}. We use the $r (Z/\sqrt{N})$ 
as our effect size measure due to its suitability for non-parametric paired data comparison. For Q2, we replicated the methodological approach used for RQ1, focusing on the correct rate per participant as our metric of interest. 

To address RQ3, we segmented the data according to question type (i.e., information retrieval and comprehension), as well as sub-types of comprehension questions (i.e., single insight versus multiple insights). We calculated the median and IQR for these segments and depicted the statistics using box plots. We then applied the same statistical tests as those used for RQ1 and RQ2 to assess differences in efficiency and effectiveness between data storytelling and conventional visualisations across varying tasks and levels of comprehension complexity. To account for multiple comparisons, we adjusted the p-values using the Bonferroni correction, multiplying the original p-values by 4 \citep{Jafari2019}. 

To address RQ4, we calculated the frequency of participant preferences for data storytelling elements that may have enabled quick identification of relevant information or facilitated accurate identification of insights. Additionally, we conducted a thematic analysis of participant explanations for the perceived utility of data storytelling elements, coding each of the 102 responses \cite{Braun_Clarke_2012}. One researcher initially led this qualitative analysis, identifying emerging themes, while a second researcher cross-verified these themes and collaborated with the rest of the authors in finalising them, thereby ensuring both the reliability and validity of the thematic interpretation \cite{Guest_MacQueen_Namey_2012}.

To address RQ5, the relationship between VL and task performance metrics was examined across two visualisation conditions: data storytelling and conventional visualisation. Two multiple linear regression analyses were conducted. The first regression analysis investigated the association between the participants' VL  and task efficiency, measured as the average success time (\textit{AvgSuccessTime}). Additionally, an interaction term (\textit{\(\text{VLAT} \times \text{Condition[DS]}\)}) was included to assess whether the visualisation condition influenced this relationship. Similarly, the second regression analysis explored the association between participants' VL and task effectiveness, measured as the correct rate (\textit{CorrectRate}). We also included the same interaction term to evaluate the potential influence of the visualisation condition on this relationship. In both regression analyses, \textit{\(\beta_0\)} represented the intercept with the conventional visualisation condition served as the baseline, while \textit{Condition[DS]} denoted the data storytelling condition. The assumptions of homoscedasticity, normality, independence of residuals, linearity, and multicollinearity were validated for both regressions.

\begin{equation}
\text{AvgSucessTime} = \beta_0 + \beta_1 \times \text{VLAT} + \beta_2 \times \text{Condition[DS]} \\
\quad + \beta_3 \times (\text{VLAT} \times \text{Condition[DS]})
\end{equation}

\begin{equation}
\text{CorrectRate} = \beta_0 + \beta_1 \times \text{VLAT} + \beta_2 \times \text{Condition[DS]} \\
\quad + \beta_3 \times (\text{VLAT} \times \text{Condition[DS]})
\end{equation}

\section{Results}

\subsection{Preliminary Exploration}
Before examining our research questions, we conducted a preliminary analysis of participants' responses. On average, participants completed the study in 40.7 minutes (std. dev.= 8.4). To find out if comprehension questions took longer to answer than information retrieval questions, we used Mann-Whitney tests. This analysis revealed that comprehension questions took longer to answer (median: 37.497 seconds, IQR: 17.279) than information retrieval questions (median: 28.043 seconds, IQR: 13.948). This difference was statistically significant and had a moderate effect size (U = 4423, Z = 4.036, p < 0.0001, r = 0.281). This is consistent with Bloom's Taxonomy, where Level 1 questions are typically simple, and Level 2 questions demand in-depth understanding.

Further breaking down the comprehension questions, we saw that questions seeking multiple insights took more time (median: 44.572 seconds, IQR: 31.587) than those asking for a single insight (median: 28.392 seconds, IQR: 15.828) or information retrieval questions. These differences were statistically significant and both had a large effect size (U = 4833, Z = 6.85, p < 0.0001, r = 0.477; U = 4960, Z = 6.685, p < 0.0001, r = 0.466). However, there was not a notable difference in the time taken to respond to information retrieval questions compared to single insight comprehension questions. This trend confirms that questions requiring multiple insights naturally take more time to answer compared to the ones focusing on a single insight, which took roughly the same time as the information retrieval questions.

Our analysis revealed that participants correctly answered comprehension questions less frequently (median: 0.556, IQR: 0.334) compared to information retrieval questions (median: 0.778, IQR: 0.333). This difference was substantial and significant with a moderate effect size (U = 231, Z = -5.87, p < 0.0001, r = 0.409). When we separated the comprehension questions into different sub-types for a more detailed analysis, we observed that questions that sought multiple insights had a lower success rate (median: 0.556, IQR: 0.445) compared to questions that aimed for a single insight (median: 0.778, IQR: 0.444). This finding was also statistically significant and exhibited a moderate effect size (U = 523.5, Z = -4.788, p < 0.0001, r = 0.334). However, there was no notable difference in the success rates of information retrieval questions and single insight comprehension questions. This demonstrates that comprehension questions requiring multiple insights were notably more challenging than the other types, which aligns with our expectations. 

\hl{We also analysed the potential impact of changing chart types (see Table }\ref{tab:charts})\hl{ using Mann-Whitney tests. Our exploration found that the average success time difference between data storytelling and conventional visualisations with chart type changes (median: -0.087 seconds, IQR: 17.682) was not significantly different from that without chart type changes (median: -0.332 seconds, IQR: 15.128) (U = 2097, Z = -0.737, p = 0.4625, r = 0.051). Similarly, the correct rate difference between data storytelling and conventional visualisations with chart type changes (median: -0.087 seconds, IQR: 17.682) was not significantly different from that without chart type changes (median: 0.167, IQR: 0.417) (U = 2240, Z = -0.152, p = 0.8801, r = 0.011). These findings indicate that altering the chart type when converting traditional visualisations to data storytelling did not significantly affect the efficiency or effectiveness of the visualisations}.

The next subsections delve into the analysis comparing data storytelling with conventional visualisations.

\subsection{RQ1 - Data Storytelling and the \textit{Efficient} Communication of Insights}
\begin{figure*}[h!]
\centering
  \includegraphics[width=.93\textwidth]{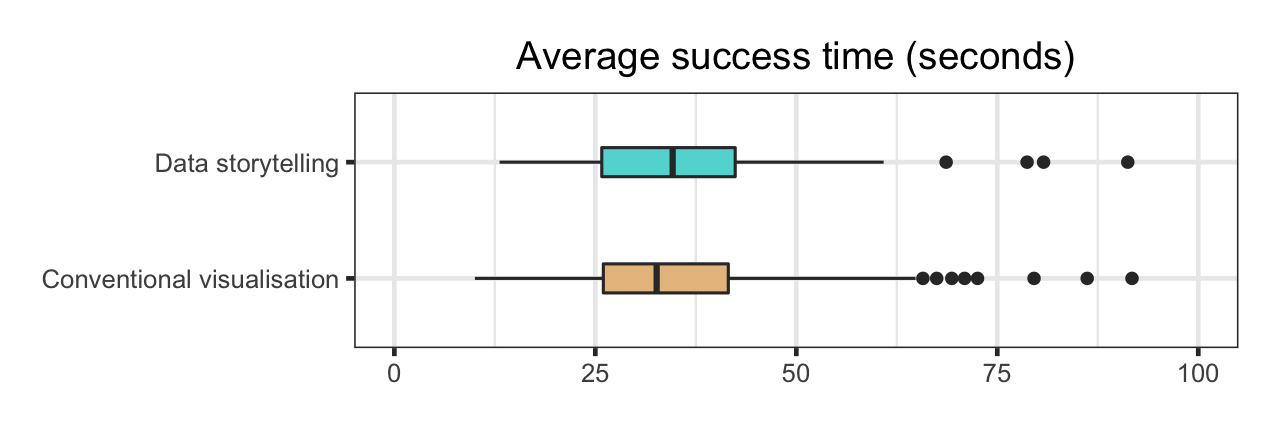}
    \setlength{\belowcaptionskip}{-5pt}
  \caption{Average success time (in seconds) of data storytelling versus conventional visualisations. The difference was not significant.} 
  \label{fig:fig6}
\end{figure*}

First, we explored whether data storytelling indeed assisted the participants in discerning and understanding critical data insights more \textit{efficiently} compared to conventional visualisations. Contrary to expectations that data storytelling could enhance the efficiency of communicating insights, our findings indicate no significant difference (U = 2505, Z = 0.195, p = 0.8465, r = 0.014) in the metric of \textit{average success time} between data storytelling (median: 34.618 seconds, IQR: 16.599) and conventional data visualisation (median: 32.625 seconds, IQR: 15.548), when considering all the tasks presented to the participants as a whole (see Figure \ref{fig:fig6}). 

These results suggest that conventional visualisation methods hold their ground in terms of efficiency, opening up an avenue for further exploration into other potential benefits of data storytelling, such as improved comprehension, which we explore next. 


\subsection{RQ2 - Data Storytelling and the \textit{Effective} Communication of Insights}

\begin{figure*}[h!]
\centering
  \includegraphics[width=.93\textwidth]{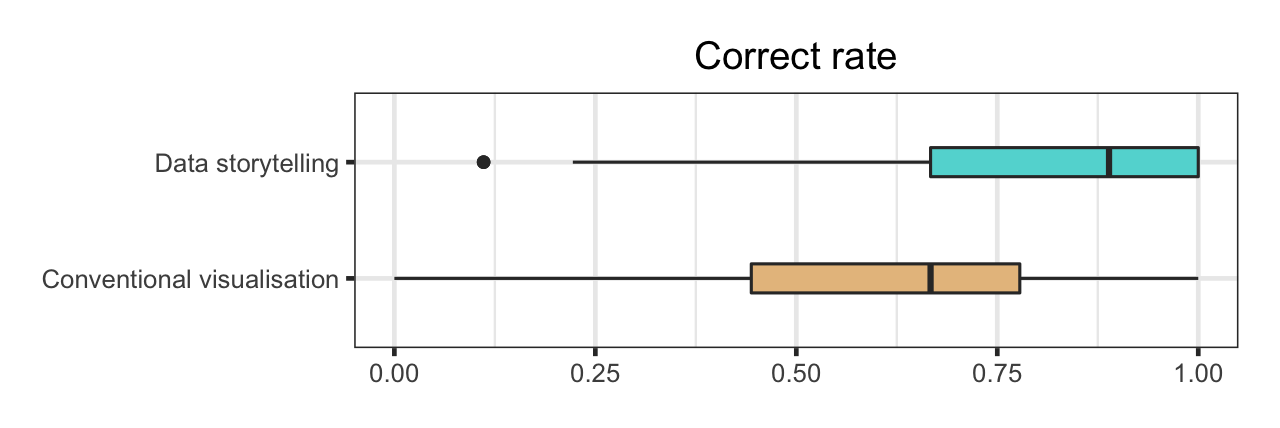}
    \setlength{\belowcaptionskip}{-5pt}
  \caption{Correct rate of data storytelling versus conventional visualisations. The difference was significant (p < 0.0001).} 
  \label{fig:fig7}
\end{figure*}

When exploring whether data storytelling assisted the participants in discerning and understanding critical data insights more \textit{effectively} compared to conventional visualisations, we discovered a significantly higher \textit{correct rate} with moderate effect size for data storytelling questions (median: 0.889, IQR: 0.333) than for conventional data visualisation questions (median: 0.667, IQR: 0.334); U = 3553, Z = 4.733, p < 0.0001, r = 0.33 (see Figure \ref{fig:fig7}). This suggests that incorporating data storytelling elements can enhance the accuracy of responses to questions pertaining to key data points and comprehension of insights when assessed across all tasks undertaken by participants. While these results underscore the potential for data storytelling to substantially improve the communication of critical data points and insights, a more detailed analysis was required, exploring variations in responses across different types of questions — including those concerning information retrieval and the comprehension of single or multiple insights. The results of further analysis are reported in the next section.


\subsection{RQ3- Data Storytelling and Information Retrieval and Comprehension Questions}
In this section, we present a more detailed analysis concerning the efficiency and effectiveness of data storytelling in comparison to conventional visualisations, taking into account the types of questions posed (information retrieval and comprehension) as well as the comprehension question sub-types (one insight and multiple insights).

\subsubsection{Efficiency and Information Retrieval and Comprehension}
We found that the \textit{average success time} for information retrieval questions was actually higher when participants were presented with data storytelling visualisations (median: 31.859 seconds, IQR: 21.166) compared to conventional visualisations (median: 26.259 seconds, IQR: 14.208) for information retrieval questions, but this difference was not significant after p-value correction (U = 3517, Z = 2.229, p = 0.1027, r = 0.155). 

Contrary to this, although the \textit{average success time} for comprehension questions (considering both sub-types together) was slightly lower with data storytelling (median: 33.461 seconds, IQR: 18.327) compared to conventional visualisations (median: 40.297 seconds, IQR: 22.341), this difference was not significant (U = 1408, Z = -2.296, p = 0.0861, r = 0.16).

\begin{figure*}[h!]
\centering
  \includegraphics[width=.93\textwidth]{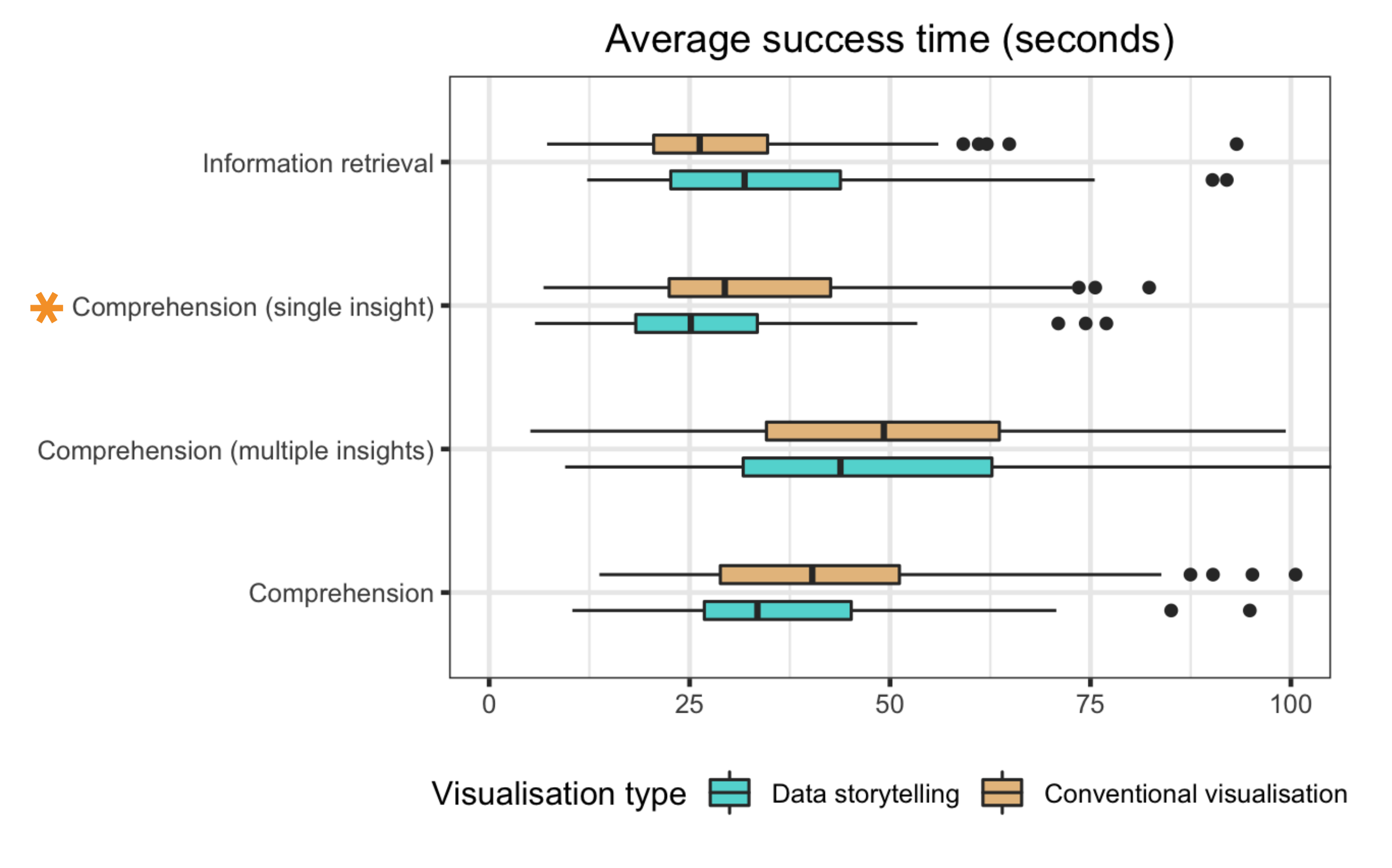}
    \setlength{\belowcaptionskip}{-5pt}
  \caption{Average success time (in seconds) of each type of question for conventional visualisations and data storytelling (* indicates the difference was significant -- p < 0.05)} 
  \label{fig:fig8}
\end{figure*}

Separating the comprehension questions further into single insight and multiple insight questions revealed that the average success time of single insight questions was lower with data storytelling visualisation (median: 25.15 seconds, IQR: 15.161) compared to conventional visualisations (median: 29.394 seconds, IQR: 20.138) and this difference was significant following p-value correction with a small effect size (U = 1192, Z = -2.686, p = 0.0283, r = 0.187). Notwithstanding, data storytelling decreased the average success time of single insight comprehension questions by an average of 4 seconds. We found that the \textit{average success time} for multiple insight comprehension questions was only slightly lower with data storytelling (median: 43.812 seconds, IQR: 31.031) compared to conventional visualisations (median: 49.232 seconds, IQR: 29.053); however, this difference was also not statistically significant (U = 1285, Z = -1.378, p = 0.6757, r = 0.096). 

In sum, conventional visualisations actually facilitated quicker responses in information retrieval tasks compared to data storytelling. Although data storytelling showed a trend towards reduced success times in comprehension tasks, particularly for questions targeting single insights, the differences were not significant.

\subsubsection{Effectiveness and Information Retrieval and Comprehension Questions}
In evaluating data storytelling as a more effective medium for communicating both critical data points and insights and facilitating comprehension, we found strong and consistent evidence in its favour for all the types of questions posed to participants. First, our findings indicated a higher \textit{correct rate} for information retrieval questions when participants utilised data storytelling (median: 1, IQR: 0.222) as opposed to conventional visualisations (median: 0.778, IQR: 0.444). This difference was significant and had a moderate effect size(U = 1716, Z = 4.157, p = 1e-04, r = 0.29), with data storytelling enhancing the correct rate by an average of 0.15. Furthermore, we noted a higher correct rate in comprehension questions using data storytelling (median: 0.778, IQR: 0.444) compared to conventional visualisations (median: 0.556, IQR: 0.556), a difference that was also significant and exhibited a moderate effect size(U = 2973.5, Z = 4.104, p = 1e-04, r = 0.286). Data storytelling increased the correct rate by a mean of 0.19, implying a more substantial improvement in comprehension questions relative to information retrieval questions.

\begin{figure*}[h!]
\centering
  \includegraphics[width=.93\textwidth]{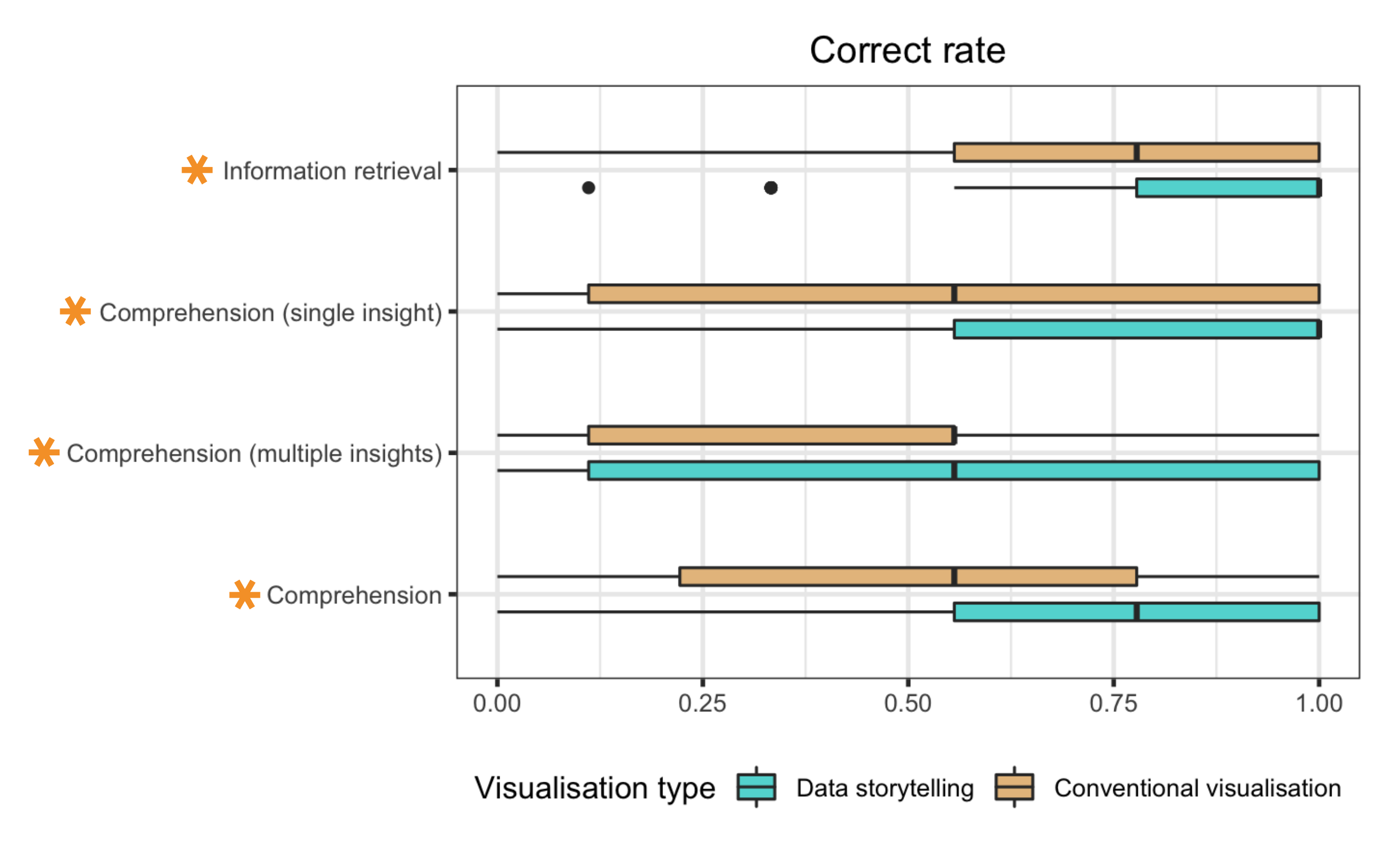}
    \setlength{\belowcaptionskip}{-5pt}
  \caption{Corrects rate of each type of question for conventional visualisations and data storytelling (* indicates significant (p < 0.05))} 
  \label{fig:fig9}
\end{figure*}

Upon dividing the comprehension questions into single and multiple insights sub-types, we observed a significantly higher correct rate for single insight questions for data storytelling (median: 1, IQR: 0.444) compared to conventional visualisations (median: 0.556, IQR: 0.889), with a significant difference and a moderate effect size (U = 1783, Z = 4.231, p < 0.0001, r = 0.295) and a mean increase in the correct rate of 0.22. Similarly, data storytelling proved more effective for multiple insights comprehension questions (median: 0.556, IQR: 0.889) compared to conventional visualisations (median: 0.556, IQR: 0.889), albeit to a lesser yet significant degree with a small effect size (U = 1613, Z = 2.667, p = 0.0304, r = 0.186), improving the correct rate by an average of 0.15. These findings illustrate a pronounced enhancement in the correct rate for single insight questions through data storytelling compared to both single and multiple insights comprehension questions.

In sum, our findings emphasise the benefits of data storytelling in enhancing both information retrieval and comprehension, especially manifest in single insight comprehension tasks, thereby underscoring its potential critical role in enabling accurate and insightful data comprehension.

\subsection{RQ4 - Perceptions on the Helpfulness of Data Storytelling Elements}

\begin{figure*}[h!]
\centering
  \includegraphics[width=.85\textwidth]{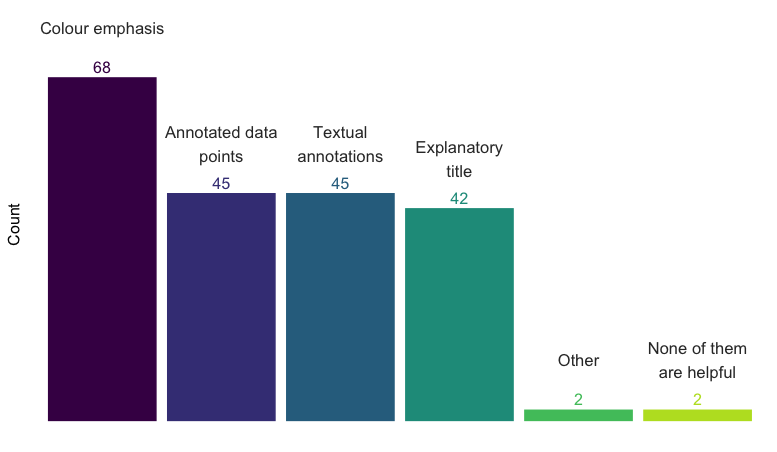}
    \setlength{\belowcaptionskip}{-5pt}
  \caption{Information retrieval: what elements helped quickly identify the relevant information?} 
  \label{fig:fig10}
\end{figure*}

\begin{figure*}[h!]
\centering
  \includegraphics[width=.85\textwidth]{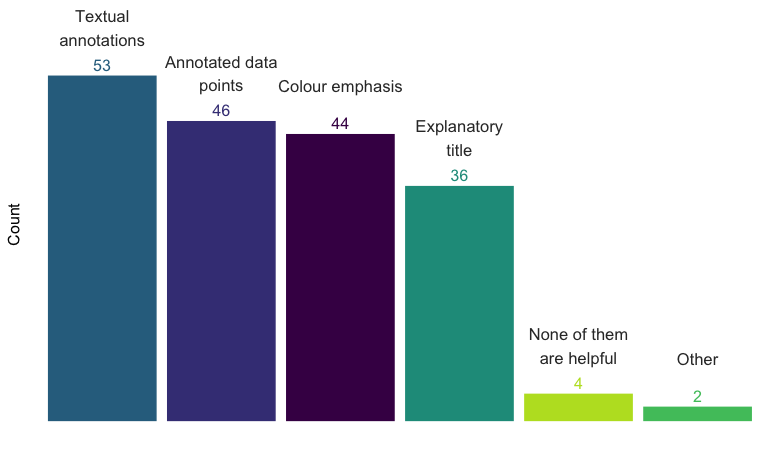}
    \setlength{\belowcaptionskip}{-5pt}
  \caption{Data comprehension: what elements helped in identifying the insights?}
  \label{fig:fig11}
\end{figure*}

We captured the participants' perceptions about how Data Storytelling (DS) elements contributed to effectiveness and efficiency. 
All the participants, but one out of 102, said that data storytelling elements were helpful for them to identify the findings quickly. Colour emphasis was regarded by more than half of the participants (68 out of 102) as helpful in increasing their efficiency in answering those questions, as depicted in Figure \ref{fig:fig10}. Textual elements, such as text annotations (N=45), annotated data points (N=45) and explanatory titles (N=45), were also perceived as contributing to identifying relevant information in the data stories quickly.
When asked which elements actually helped them respond to the questions, about half of the participants (N=52) indicated that text annotations were helpful, followed by annotated data points (N=46) and colour emphasis (N=44) (see Figure \ref{fig:fig11}).

Put together, these results suggest that the participants perceived the colour emphasis as helpful to drive attention to the critical data points of the data story, but the textual information in the form of annotations was considered critical to gain an understanding of the insights. Below, we elaborated on five themes that emerged when participants were asked to further explain their responses. 

\subsubsection{Theme 1 - Perceived DS usefulness to support data understanding.}
The first theme referred to the usefulness of data storytelling visual elements to \textbf{\textit{identify key information to support the understanding of the data and reduce complexity}} (N=24). Visual elements like \textit{"added brief explanations"} [textual annotations] (P15), \textit{"colour schemes" } [colour emphasis](P26, P29), and \textit{"clear labelling"} [annotated data points] (P27) helped participants to \textit{"identify the important elements"} on the graph more readily (P15, P18, P22). The participants emphasised the role of textual elements, whether annotations or explanatory titles, in making the data \textit{"easier to understand"} (P66, P67). These storytelling elements were not merely facilitative but often essential for understanding the graph: (\textit{"some graphics are harder to interpret, and the text} [annotations]\textit{ helped me to understand them [graphics] better"} -- P49), as they \textit{"helped to give facts rather than estimations"} (P19, P36). 
 
\subsubsection{Theme 2 - More \textit{efficient} interpretation of data.}
Another theme that emerged referred to the perceived role of data storytelling elements in enhancing \textit{\textbf{their efficiency when identifying key information}}  (N=26).
For example, some participants explained that these elements acted as guides that helped them \textit{"identify more efficiently the relevant information"} (P60, P67, P68) and \textit{"[convey] the main idea and the key points faster"} (P67), sometimes even obviating the need for detailed chart analysis: "\textit{Without analysing the chart I would get the information needed just by reading the textual annotations"} -- P61. 
This efficiency could be attributed to the use of data storytelling elements to focus their visual attention. Textual elements combined with colour emphasis helped the participants to \textit{"focus [their] attention"} (P70) and find answers \textit{"way quicker"} (P65). One participant elaborated on this as follows: \textit{"I tend to get distracted if data is only represented by numbers or plain text. Colours and Explanatory titles were really helpful"} -- P56.

Colour emphasis also played a role in helping participants \textit{"find answers way quicker"} (P65) and making the data \textit{"stand out"} (P69), effectively directing their attention towards accurate information: \textit{"without them, I would not know what I am looking at."} -- P43; \textit{"it points to the correct data analysis; without it, you cannot tell what it is about or stands for"} -- P8. 
Participants also noted that helping them to focus their attention on what was important in a visualisation was particularly beneficial when operating under time constraints: \textit{"breaking down the data with colour seems to help; it drew my attention, especially when you're being timed"} -- P44. These findings highlight the perceived role of data storytelling elements in focusing their attention and improving their efficiency when identifying key information. 

Participants (N=10) also emphasised the importance of data storytelling elements in \textit{\textbf{enhancing the readability and depth of understanding associated with data visualisations}}. One participant (P2) noted that \textit{"short text indicating values or indications enhances the readability"}, highlighting the role of textual elements in making the graphs more accessible. The perceptions of the contributions of the added context provided by data storytelling elements in enabling quicker data comprehension were expressed by another participant (P6): \textit{"It helped me to understand the data more quickly, and it gave me more context of it"}. Participants confirmed that the textual elements contributed to a more holistic understanding of the data; they \textit{"give additional information to the graph"} (P7); \textit{"give a detailed look on the data"} (P3) and help \textit{"give facts rather than estimations"} (P12). Moreover, the storytelling elements added a layer of depth, making the information retrieved \textit{"much more complete"} (P14). These comments suggest that storytelling elements can improve the ease of data interpretation and serve as complementary information to make the data more accessible and contextual by highlighting relevant information.

\subsubsection{Theme 3 - Easier and more accurate interpretations of the data.}
Participants (N=21) also perceived that the storytelling elements enhanced their ability to \textit{\textbf{quickly and accurately interpret the data}}. 
 Colour emphasis and annotated data points were acknowledged for their role in making data \textit{"easier to interpret"} (P42, P43, P47) and supporting \textit{"more accurate"} interpretation (P51). Participants noted that explanatory titles, \hl{textual annotations} and colour emphasis made it\textit{ "easier to locate information"} (P26), enabling them to answer questions with greater confidence and accuracy (\textit{"with the storytelling elements was easier find the information to respond the questions correctly"} --P24; \textit{"it made my interpretations more accurate"} --P51). This was echoed by P16, who felt that such elements present the information at a glance to \textit{"speed up the process"} of reaching the correct answer. 
 
Similarly, \hl{textual annotations} and annotated data points were perceived as helpful to \textit{"reinforce the graphic information"} (P41) and \textit{"confirming assumptions"} (P40), thereby allowing for more accurate responses to questions. 
Interestingly, while some participants (P26, P37) felt that they could have eventually figured out their responses without these storytelling elements, they acknowledged that data storytelling elements made the identification of key information faster and more focused. This perspective was echoed by P41, who felt that data storytelling \textit{"reinforces the graphic information and helps to understand better and integrate any data that may not appear immediate at first analysis"}. These perspectives suggest that data storytelling elements make the data interpretation process more precise, allowing participants to validate their initial interpretations and derive insights with greater confidence. 

\subsubsection{Theme 4 - Negative perceptions and concerns.}
In contrast to the perceived benefits of data storytelling elements highlighted above, some participants (N=16) found that these elements were \textit{\textbf{not always effective for interpreting data visualisations}}. Such elements, particularly textual elements \hl{(i.e., annotated data points, textual annotations and explanatory titles)}, could introduce confusion (\textit{"actually made the data more confusing to look at"} --P85, \textit{"it's just a lot of information and it can be a bit overwhelming"} -- P97) or serve as distractions if not used effectively ("\textit{the highlights \hl{[textual annotations]} are kind of distracting from other data if done for only one thing}" --P87, "\textit{the \hl{[textual annotations]} distracted me from the correct visualisation of the data.}" --P100). 
Some participants (N=6) also questioned the need for these visual elements, arguing that the data was self-explanatory and that storytelling elements did not add significant value (\textit{"The data pretty much explains itself ... the ones with data storytelling did not make that much difference."} --P86) or that the non-storytelling graph was \textit{"easier to interpret"}, \textit{"clear"} and \textit{"more visually appealing"} (P88, P92, P98).
Some participants (N=3) expressed concerns about the potential for these elements to be misleading or distracting. P90 elaborated on this as follows: \textit{"I preferred to base my responses on graphs because I didn't know if the information about the visualisations \hl{[textual annotations]} were true to the presented data"}, while another found the highlights \textit{"distracting from the correct visualisation of the data"} (P100). These comments suggest that while data storytelling elements could be beneficial, these should be carefully crafted to avoid information overload, particularly when the data is easier to interpret.

\subsection{RQ5 - Visualisation Literacy}

For task efficiency, the regression model (2) did not explain a significant amount of variance, \( R^2 = 0 \), \( F(3, 534) = 0.08 \), \( p = .97 \), with only a significant intercept (\( \beta_0 = 33 \), \( t(534) = 10 \), \( p < .001 \)), indicating an average success time of 33.11 seconds when all independent variables were zero. However, all independent variables including visualisation literacy (\( \beta_1 = 0.36 \), \( t(534) = 0.06 \), \( p = .95 \)), data storytelling condition (\( \beta_2 = 1.7 \), \( t(534) = 0.39 \), \( p = .70 \)), and the interaction term (\textit{\(\text{VLAT} \times \text{Condition[DS]}\)}; \( \beta_3 = -3.2 \), \( t(534) = -0.38 \), \( p = .71 \)) were not statistically significant. 

In terms of task effectiveness, the regression model (3) explained a significant amount of variance, \( R^2 = 0.17 \), \( F(3, 560) = 38 \), \( p < .001 \). The intercept was significant (\( \beta = 0.21 \), \( t(560) = 3.3 \), \( p = .001 \)), suggesting a correct rate of 21\% when all independent variables were zero. Additionally, the participants' visualisation literacy was significant and positively related to their correct rate (\( \beta = 0.76 \), \( t(560) = 6.4 \), \( p < .001 \)), indicating that for each percentage of increase in visualisation literacy, the participants' correct rate would increase by 0.76\%. However, the data storytelling condition (\( \beta = 0.15 \), \( t(560) = 1.7 \), \( p = .096 \)) and the interaction term (\textit{\(\text{VLAT} \times \text{Condition[DS]}\)}; \( \beta = 0.016 \), \( t(560) = 0.096 \), \( p = .92 \)) were not statistically significant, indicating that the relationship between the participants' visualisation literacy and task effectiveness remained the same for both data storytelling and conventional visualisation conditions.

\section{Discussion}
In this section, we summarise the key findings, share our critical reflections on the implications for research and practice, and note the limitations of our study and potential avenues of future work. 

\subsection{Summary of Results and Research Questions}
In Dykes' \citep{dykes2015data} view, the purpose of a data story is to guide the audience to a better understanding and appreciation of an important data point or insight. Our study aimed to test whether data storytelling can assist individuals in more effectively and efficiently discerning and understanding critical data points and insights compared to conventional visualisations. We assessed the identification of critical data points through information retrieval questions and insight recognition through comprehension questions. 

\hl{Regarding RQ1, contrary to prior assumptions }\citep{Ryan2016, zhang2018converging, zhang2019designing, Zhang2022framework}, \hl{our results demonstrated no evidence that data storytelling improved \textit{efficiency} (measured by the time taken to successfully complete tasks) when evaluating both information retrieval and comprehension questions collectively. Although the existing literature suggested that data storytelling would enable more \textit{efficient identification of insights}, this assumption had not been empirically tested before our study}.
In contrast, for RQ2, strong evidence emerged suggesting that data storytelling improved \textit{effectiveness} (correct response rate) in aiding participants to more accurately comprehend insights derived from data stories compared to conventional visualisations. \hl{This finding confirms earlier non-empirical assumptions} \citep{Kosara2013,gershon2001storytelling} \hl{and the foundational definition of data storytelling by} \citet{dykes2015data}. 

Regarding RQ3, we found varied outcomes on data storytelling's efficiency for different question types. Despite predictions grounded in RQ1 outcomes, data storytelling expedited answers to single-insight comprehension questions but not to multiple insights queries. This might be attributed to the inherently greater complexity of multi-insight questions, as highlighted in our preliminary analysis. Thus, data storytelling seems particularly adept at enhancing the rapid comprehension of visualisations conveying a single central insight, aligning with existing literature that suggests that a data story should focus on a core insight or call to action (Principles \#1 \& \#5) \citep{bach2018narrative,Segel2010,Lee2015more,dykes2015data,ojo2018patterns}.
However, despite the observed difference in response times between single and multiple insight questions, our analysis revealed that data storytelling fostered more accurate answers across all question types, including both information retrieval and comprehension questions. This suggests that data storytelling effectively directs viewers to critical data points and insights within a visualisation, improving accuracy even for tasks demanding identification of multiple data points, albeit taking a longer time to complete.

Regarding RQ4, a significant number of participants (N=26) found data storytelling elements, especially coloured emphasis, beneficial in \textit{efficiently} pinpointing vital details in the graphs (Principle \#4). Furthermore, while a majority found some storytelling elements beneficial to some degree in \textit{effectively} answering the questions (i.e., text annotations and annotated data points), a smaller group (N=16) found them distracting and unhelpful, underlining the importance of judicious use of these elements in accordance with our third design principle — \textit{discerning employment of storytelling elements}. This reflects previous research suggesting that, for some, these features might be more obstructive than facilitative \citep{kong2019understanding}. The feedback also hinted at a possible adverse effect of using data storytelling for individuals proficient in data analysis, with some preferring conventional visualisations with detailed data over textual annotations \citep{kalyuga2009expertise}.

Regarding RQ5, we investigated how an individual's visualization literacy impacts the efficacy of data storytelling. Our findings indicated a moderate to strong relationship between visualisation literacy and the accuracy rates for both data storytelling and conventional visualisation. Contrary to popular belief upheld in prior studies \citep{Ma2012,Zhang2022framework,Figueiras2014}, our results did not statistically affirm that data storytelling is more beneficial for individuals with lower visualisation literacy. This implies that data storytelling is uniformly advantageous for users, regardless of their visualisation literacy level.

\hl{In summary, based on our results, we provide the following suggestions to design practitioners interested in using data storytelling elements for communicating insights: }
\begin{enumerate} 
    \item \hl{Data stories can be more effective compared to conventional visualisation in communicating insights, although the data stories may not necessarily help in finding the insights faster} (RQ1 \& RQ2).
    \item \hl{Data stories may be more efficient compared to conventional visualisations for tasks that involve identifying a single insight but not necessarily for searching specific data points or identifying multiple/complex insights from a visualisation. (RQ3) }
    \item \hl{For information retrieval tasks, especially for users with strong analysis expertise, it may be preferable to reconsider the need for adding data storytelling elements to conventional visualisations. (RQ4)}
    \item \hl{Excessive addition of data storytelling elements, particularly those that are text-based, may hinder simple data tasks or be distracting for certain users. (RQ4)}
    \item \hl{Data stories aid in the effective identification of insights relatively equally for all users, regardless of their data literacy levels. (RQ5)}
    \item \hl{At the same time, a higher level of data literacy in a user may lead to more effective identification of insights, whether using a data story or a conventional visualisation. (RQ5)}
\end{enumerate}


\subsection{Implications for Research}
\hl{While previous research has yielded mixed results regarding the benefits of data storytelling in areas such as engagement} \citep{boy2015does,zhao2019understanding}, \hl{empathy} \citep{boy2017showing,morais2021can,Liem2020}, \hl{and memory recall} \citep{Zdanovic2022}, \hl{our study supports the fundamental notion that data storytelling is an effective medium for conveying insights. Our findings add to this body of work by demonstrating that data storytelling not only effectively aids in identifying insights but also does so more efficiently when pinpointing a single insight within a visualisation. However, this effectiveness and efficiency in insight identification does not necessarily enhance memory recall, as previously suggested by }\citet{Zdanovic2022}. \hl{Moreover, even if the insights can be identified more effectively, adding annotations and other storytelling elements to conventional visualisations may not always increase engagement }\citep{boy2015does,zhao2019understanding} \hl{or empathy towards the subject matter} \citep{boy2017showing,morais2021can,Liem2020}. \hl{Our results did not directly contradict earlier negative studies, as these did not measure the effectiveness and efficiency of data storytelling in information retrieval and comprehension of insights. Consequently, our study represents an initial step in a broader research agenda, advocating for more comprehensive investigations into the role and limits of data storytelling, a tool increasingly prevalent across various sectors} \citep{sun2023data,park2023charlie,yang2023pair,schroeder2020evaluation}.

Our results resonate with the notion posited by \citet{dykes2015data} and others, portraying data storytelling as an \textit{effective} tool in conveying insights \citep{Segel2010,Knaflic2015,gershon2001storytelling,krum2013cool,Daradkeh2021}. Our results hold substantial implications for research, highlighting the key role data storytelling can play in amplifying data comprehension and communication. Researchers in fields such as data science, journalism, and education can potentially explore how to adopt its foundational design principles for broader audience engagement \cite{bottinger2020reflections,lee2015people}. 

Moreover, they underpin the necessity for further studies to delve into the optimal ways of employing data storytelling elements, as several different variations of data storytelling exist  \citep{bach2018narrative}, aiming not just for clarity and effectiveness of insight communication but also fostering a deeper connection with the audience. Furthermore, the critical feedback received in our study signals a pathway for investigating the potential downsides of data storytelling, for example, considering the ethical considerations of emphasising certain stories over others, which has started to be explored by \citet{fernandez2021storytelling}.

Moreover, our research also sheds new light regarding the suitability of data storytelling to particularly support general audiences with limited data visualisation expertise, presenting a different picture compared to the findings of \citet{Pozdniakov2023ectel}. Despite their smaller study highlighting a reduced cognitive load among 23 teachers through the use of basic data storytelling elements, we did not find a direct link between visualisation literacy and data storytelling effectiveness. However, we noted that visualisation literacy did affect the ability to discern insights in both conventional visualisations and data stories. This opens avenues for deeper exploration into the intricate relationship between visualisation literacy and data storytelling, encouraging a broader understanding of the role visualisation literacy plays in the realm of data visualisation.


\subsection{Implications for Practice}

This study validates emerging guidelines for practitioners in various industries, confirming the potential of data storytelling to enhance information retrieval and comprehension tasks \citep{Knaflic2015,feigenbaum2020data,ryan2018visual}. It also illustrates that data storytelling can be a vital tool for marketing campaigns and educational purposes, where precise data comprehension is paramount, proving especially efficient for single-insight tasks \citep{dykes2015data, martinez2020data, wang2019teaching, shan2022heritage}. The proven effectiveness (and efficiency for single insight tasks) can be invaluable in decision-making contexts, highlighting its role in facilitating both timely and accurate interpretation of data representations \citep{Daradkeh2021, schroeder2020evaluation}.

Interestingly, we observed a slightly prolonged response time in the data storytelling condition for information retrieval questions, although the difference was not significant. This finding prompts a cautionary note for designers to avoid overloading data stories with excessive elements such as too detailed annotation texts and lengthy explanatory titles, which can potentially hinder simple tasks \citep{tufte2001visual, Knaflic2015, feigenbaum2020data, ryan2018visual}. Thus, presenters must judiciously determine the cognitive demands of their tasks and tailor their use of data storytelling to meet those \citep{bach2018narrative, Segel2010, Lee2015more, dykes2015data}.

Our study underscored that data storytelling aids users of all visualisation literacy levels equally, hinting at its potential utility compared to conventional data visualisations, especially for the general public. This approach, already adopted successfully by newspapers, could potentially revolutionise the way research findings are communicated through scientific visualisations, making them more accessible to the public \citep{ojo2018patterns,Ma2012}. Leveraging data storytelling in places like museums and scientific events could enhance the comprehension of the complex data being presented, encouraging a deeper understanding \citep{Borner2016}.


Looking ahead, there exists a promising avenue for creating automated or AI tools to seamlessly blend data storytelling elements into data visualisations, a frontier with considerable room for growth, despite existing progress in automatic annotation creation \citep{li2023ai, Wilkerson2021youth, fernandez2021storytelling, lai2020automatic}.

\subsection{Limitations and Future Work}
Our study has several acknowledged limitations\hl{ in terms of the data collection and the study design decisions. Regarding \textit{data collection}}, the participant pool, constrained by the recruitment platform and available resources, lacked ideal diversity with an over-representation from the Americas and an under-representation of other regions. Moreover, we restricted our exploration to six pairs of visualisations, limiting the breadth of data storytelling elements and visualisation types assessed. Future research should aim to involve a broader, more diverse participant group and extend the range of visualisations studied to foster a more comprehensive analysis \citep{Pouliquen2004languages,Lee2017}.

Moreover, the study was conducted exclusively in English, and while participants were required to be proficient in English, their perceived proficiency levels were not documented. Given that data storytelling substantially emphasises narrative -- which can inherently vary across languages -- our results might not hold the same ground in different linguistic contexts \citep{Pouliquen2004languages}. Therefore, further work is recommended to reproduce this study in other languages. Additionally, the online format of the study potentially encouraged hurried participation due to the fixed monetary compensation. Moreover, concurrent distractions could have further diluted participants' focus. Future endeavours could look to include deeper qualitative analysis through offline methodologies such as interviews to avoid these issues.
\hl{In the same vein, we summarised the overall perceptions of data storytelling elements in terms of their effectiveness and efficiency in identifying insights, based on responses to questions. Future studies could utilise eye-tracking technology to investigate how each data storytelling element contributes to the efficiency and effectiveness of retrieval or comprehension tasks, similar to smaller-scale studies like those in} \cite{Echeverria2018} and \cite{Pozdniakov2023how}. 

\hl{Regarding the \textit{study design}}, we tailored the visualisation literacy test (VLAT) to focus specifically on the visualisations central to our study, sidelining other pertinent techniques featured in the complete VLAT \citep{Lee2017}. This strategy, while facilitating the study execution, highlights the necessity for future research to investigate the visualisation techniques that were not considered in our study. Moreover, the scope of our inquiry centred predominantly on the initial two levels of Bloom's taxonomy, limiting the study to evaluating data storytelling's aid in key data points retrieval and insights comprehension. Subsequent studies can delve into higher tiers of the taxonomy to evaluate aspects such as data story manipulation and the creation of personal data stories \citep{arneson2018visual,Burns2020,mnguni2016assessment,Byrd20197d}.

\hl{Additionally, our study design considered changes in the type of charts when transforming from conventional visualisations to data stories, adhering to recommendations from the literature} \cite{Knaflic2015,Zdanovic2022}. \hl{ The results revealed no significant differences when changing the type of chart. Yet, further studies may consider a deeper exploration of how this change could impact efficiency and effectiveness by considering the participants' spatial skills } \cite{Tandon2023}. \hl{Moreover, although we followed  design guidelines by }\citet{Zdanovic2022} and \citet{Knaflic2015} \hl{to transform conventional visualisations into data stories, and the process was jointly performed by three researchers, external expert validation can be recommended to provide additional credibility to the data storytelling construction process. }
Lastly, we limited our investigation to static data storytelling to maintain study simplicity, excluding interactive elements that can enrich data narratives. Future work can benefit significantly from exploring interactive patterns in data storytelling to enhance user engagement \citep{bach2018narrative}.

\subsection{Conclusion}
As society grapples with the overwhelming volume of data generated every moment, data storytelling emerges as a potent tool promising to distill complex data into discernible insights, thereby facilitating informed decision-making. Our research shows that data storytelling can be a way to identify key data points and insights more accurately and effectively, even though it might not always be faster than using conventional visualisations.
We also noticed that being skilled at understanding visualisations is key to getting the most out of them, but data storytelling is helpful for people at all visualisation literacy levels. 
We argue, supported by our study, that this work advances the state of the art in understanding the benefits of data storytelling for the retrieval and comprehension of key data points and insights hidden in large sets.

\begin{acks}

Roberto Martinez-Maldonado's research was partly covered by the Jacobs Foundation.

\end{acks}

\bibliographystyle{ACM-Reference-Format}
\bibliography{cas-refs}

\appendix

\end{document}